\documentclass[sigconf,authorversion,nonacm,10pt]{acmart}

\AtBeginDocument{%
  }

\setcopyright{acmlicensed}
\copyrightyear{2018}
\acmYear{2018}
\acmDOI{XXXXXXX.XXXXXXX}

\acmConference[Conference acronym 'XX]{Make sure to enter the correct
  conference title from your rights confirmation emai}{June 03--05,
  2018}{Woodstock, NY}
\acmISBN{978-1-4503-XXXX-X/18/06}
\usepackage{subcaption}
\usepackage[ruled,vlined]{algorithm2e}
\usepackage{wrapfig}
\usepackage{url}

\begin{document}

\title{Near-realtime Earth Observation Via LEO Satellite Constellation}

\author{Bo Wu}
\email{bow36@pitt.edu}
\affiliation{%
  \institution{University of Pittsburgh}
  \city{Pittsburgh}
  \state{PA}
  \country{USA}
}

\author{David Tipper}
\email{dtipper@pitt.edu}
\affiliation{%
  \institution{University of Pittsburgh}
  \city{Pittsburgh}
  \state{PA}
  \country{USA}
}

\author{Pengfei Zhou}
\email{pengfeizhou@pitt.edu}
\affiliation{%
  \institution{University of Pittsburgh}
  \city{Pittsburgh}
  \state{PA}
  \country{USA}
}


\begin{abstract}
Earth observation (EO) satellites in Low Earth Orbit (LEO) are collecting vast amounts of invaluable data for applications such as monitoring forest fires. However, data downloading from EO satellites faces significant challenges due to the limited number of ground stations and the brief communication windows with them. Conversely, emerging LEO constellations like Starlink have enabled continuous connectivity and revolutionized access for ordinary users globally. In this paper, we study the feasibility of supporting EO satellites with Starlink satellite infrastructure and introduce a novel data delivery system, designated as “Starlink Space User” (SSU), for relaying data from observation satellites. SSU treats EO satellites as space users of Starlink. At the core of SSU is a novel class of algorithms designed for link and Internet Point of Presence (PoP) selection, as well as system scheduling optimization, that operate effectively atop Starlink’s proprietary infrastructure. We assess the performance of SSU using trace-driven evaluations alongside real-world Starlink performance measurements. Our results demonstrate that the proposed Starlink-aided design can significantly reduce the median backlog (data not delivered) per satellite.
\end{abstract}

\maketitle


\section{INTRODUCTION}
The use of Earth observation (EO) satellites in Low Earth Orbit (LEO) has become increasingly vital for a wide range of applications, including climate monitoring \cite{Weather-forecasting}, disaster response \cite{WOOSTER2021112694}, and agricultural management \cite{LEOforagriculture}. Today’s EO systems operate as constellations comprising hundreds of small satellites, which typically orbit at altitudes of 370$\sim$430 km \cite{UCSSatelliteDatabase} and can generate data at an extraordinary scale, up to several terabytes per satellite per day \cite{L2D2, Colton2016SupportingTF}. While these data deliver insights with unprecedented accuracy and immediacy, the process of transferring them to ground stations is hindered by significant logistical challenges. LEO satellites are traveling in low earth orbit and appear extremely fast to ground stations where the data should be transferred. They can typically only contact a ground station in four to six ten-minute windows per day per satellite-ground station pair \cite{Devaraj_Ligon_Blossom_Breu_Klofas_Colton_Kingsbury_2019}. 
Due to the high data volume and limited bandwidth, there are large inherent delays in transferring the data, which can be day-level \cite{L2D2} and critically impact the timeliness and usefulness of the data, particularly in scenarios where real-time information is paramount, such as emergency response during natural disasters.


Current efforts to address these challenges focus on improving existing ground stations or optimizing communication protocols. Strategies such as increasing the downlink rate during these limited contact times \cite{10.1145/3570361.3592521, 10.1145/3636534.3649362}, deploying additional small ground stations \cite{L2D2, 10.1145/3422604.3425926}, and refining data querying processes \cite{295557} have been studied. While these solutions provide partial relief, they often involve substantial investments in infrastructure or still rely on outdated communication frameworks, missing out on leveraging advancements in modern satellite communication technologies.


Inspired by the rapid expansion of LEO satellite communication networks like Starlink \cite{Starlink}, this paper explores the potential for enabling data download from EO satellites via LEO satellite constellation. As of April 2025, SpaceX \cite{SpaceX} has more than 7061 satellites in orbit mostly at the altitude of 550 km \cite{mobicom23li, nsdi24liu}. While this infrastructure is designed to serve terrestrial users, it also presents a promising opportunity to accelerate data transfer for space users like EO satellites via Ka-band links.
To assess feasibility, we conduct a preliminary analysis using real-world orbital data from over 7,000 active Starlink satellites and representative EO satellites deployed in space. Our findings indicate that selecting the nearest Starlink satellite minimizes distance but can result in high relative velocities (up to 14.5 km/s) and significant relative velocity variations, which degrade communication performance.
Instead, by slightly expanding the search radius, we found that stable low-velocity links (e.g., $\sim$7.4 km/s) are achievable in most cases. 
Moreover, in orbits close to the poles, EO satellites sometimes can maintain even lower relative velocities due to orbital alignment with Starlink’s polar planes. These results suggest that utilizing Starlink for EO data download is feasible with proper link selection strategies.

\begin{figure}
  \centering
  \includegraphics[width=0.8\linewidth]{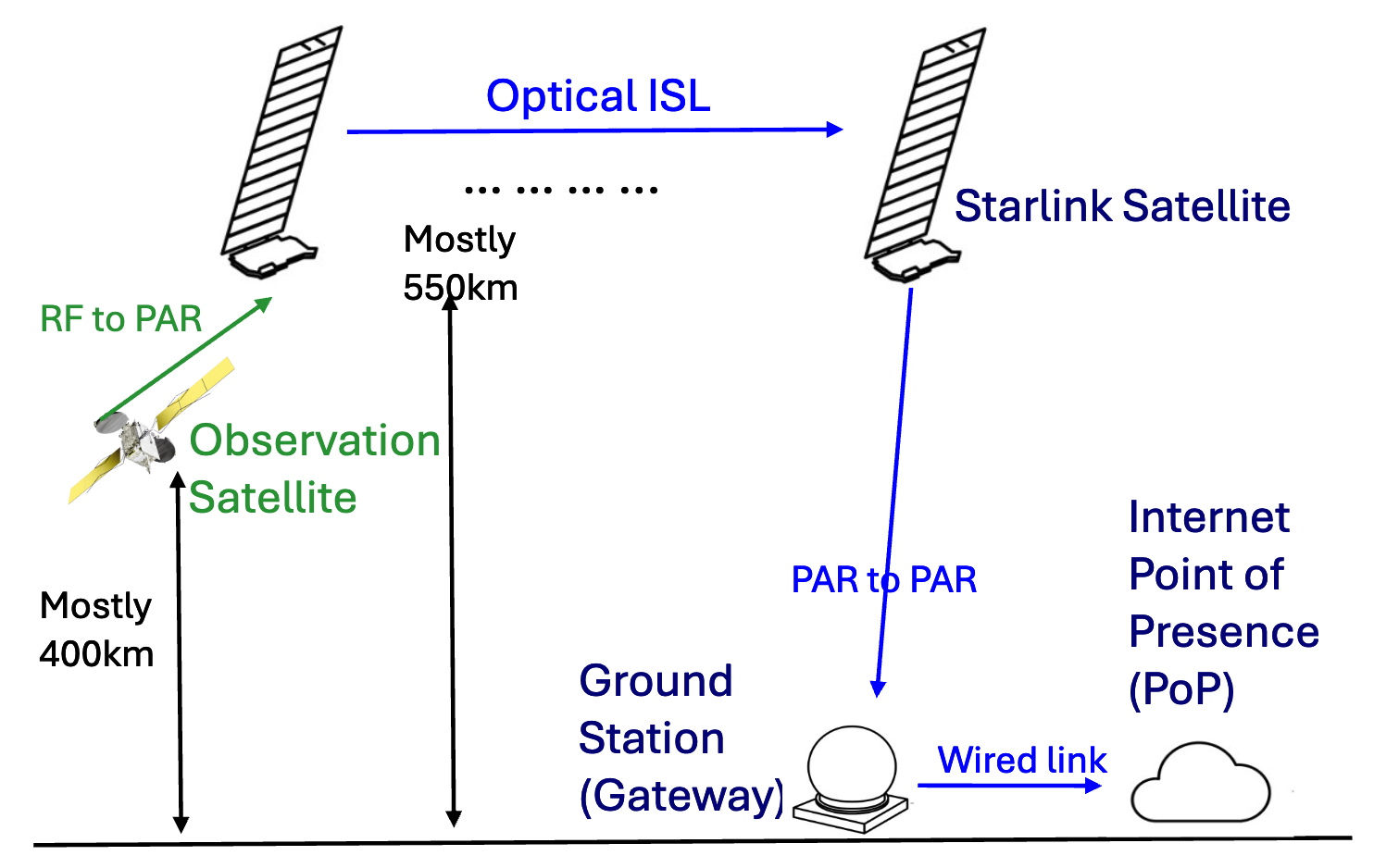}
  \vspace{-0.1in}
  \caption{SSU is a Starlink aided data transfer system.}
  \vspace{-0.25in}
  \label{fig:ssu}
\end{figure}

In this context, as illustrated in Figure \ref{fig:ssu}, we introduce "Starlink Space User" (SSU), a system that assimilates Earth observation satellites into the Starlink network, treating them as additional space users. 
The EO satellite connects to a  Starlink satellite, which relays the data through the network via Inter-Satellite Links (ISLs) to the ground station linked to an Internet point of presence (POP).
Our design builds on the existing Starlink system, aiming to preserve its core mechanisms while addressing the unique challenges posed by space users. Three key challenges are as follows.

\noindent$\bullet$\hspace{0.5em} \textbf{Dynamic destination selection:} 
The key difference between space and ground users lies in how they connect to Starlink infrastructure. Ground users are typically assigned a fixed PoP, with changes occurring rarely and usually through manual intervention due to the absence of dynamic PoP switching. In contrast, space users experience highly dynamic PoP connections as they orbit the Earth, requiring continuous rerouting of data through different PoPs. This necessitates strategic PoP selection and consideration of ISLs to optimize connectivity.
Therefore, understanding Starlink’s infrastructure is crucial: How do PoPs perform under varying conditions? How does ISL performance change with different inter-satellite distances? Although Starlink’s system is proprietary, we adopt an end-to-end data-driven approach to characterize ISLs and evaluate PoP performance using extensive ground user communication data.



\noindent$\bullet$\hspace{0.5em} \textbf{Link selection strategy:}
The second challenge for space users is selecting the appropriate Starlink satellites for connection. Using real-world Two-Line Element (TLE) data \cite{space-track}, we analyzed the distance and relative velocity between EO and Starlink satellites. While space users often benefit from shorter distances than ground users, they face much higher relative velocities, causing severe Doppler effects and link instability. To mitigate this, we developed an orbit-aware, dual-criterion selection strategy that dynamically balances proximity and frequency shift. The strategy also accounts for whether the EO satellite is inside or outside Starlink’s Service Density Boundaries (SSDB), ensuring robust and efficient link establishment across varied orbital conditions


\noindent$\bullet$\hspace{0.5em} \textbf{Systematic downlink scheduling:} 
The third challenge involves coordinating data downloads between EO satellites and Starlink at the system level. Although individual link quality can be estimated using prior models, performance varies with orbital shifts and may suffer from contention due to rising demand from both EO and ground users. To tackle this, we frame scheduling as a dynamic bipartite matching problem under link capacity constraints. We propose a throughput-optimized greedy algorithm that incorporates switching penalties to minimize handover-related performance loss \cite{Kapoor2024-vc}. This enables adaptive, high-efficiency link planning across thousands of connection opportunities.

To evaluate SSU, we use the real-world communication data for approximately 6 months from 13 Starlink dishes located across 7 PoPs on three continents: Africa, North America, and Europe \cite{zhao2024lens}. Additionally, we gathered publicly available TLE data, global PoP distribution data for Starlink, and weather-related data for all Starlink dishes included in our dataset. We use Starlink communication data without ISL, reflecting a bent pipe structure, to profile the PoPs' performance. Building on the POP profiles, we then develop our ISL estimation module using ISL-enabled communication data. 
To evaluate system scheduling, we use publicly available orbital information for Starlink and Planet’s published, frequently updated orbital data \cite{Celestrak, Planet} to assess system performance when each observation satellite has 500 GB of data to download per day \cite{295557}. 

The results are as follows. 
\textbf{(1) PoP profiling and ISL quality estimation:} SSU's PoP profiling achieves a median absolute error of \textit{0.52 \%} in predicting the route downlink packet loss rate and \textit{10.1 ms} in packet delay from a specific position on Earth to a PoP via the bent-pipe structure. This translates to an median error of \textit{39.9Kb/s} for the downlink route on standard Starlink service.
\textbf{(2) Link selection efficiency:} The link selection strategy achieves nearly 2× the data transfer efficiency compared to the baselines in our experiments.
\textbf{(3) Route quality estimation and data transfer:} SSU estimates a median route packet loss rate at \textit{0.98\%}. This allows one observation satellite to transfer \textit{63.3\%/314.0G} of its collected data to Earth in one day, which is \textit{3.4$\times$} compared to the baseline method using the traditional wait-and-transfer approach. 
When each satellite collects \textit{500 GB} of data per day, SSU reduces the median backlog (data not delivered) per satellite from \textit{407.6 GB} to \textit{186.0 GB}.

We believe that an integrated space and terrestrial communication system represents the future of communication systems. 
In summary, we make the following contributions:

\begin{itemize}
    \item We present a new space data transfer framework that leverages Starlink as infrastructure to facilitate data transfer from observation satellites and validated its feasibility by in-depth orbital analysis.
    \item We propose three novel designs, i.e., Starlink PoP profiling, orbital analysis based link selection, and system scheduling optimization, to cope with the challenges of space users and realize the system development of SSU.
    \item We evaluate the design of SSU by extensive experiments using real-world measurements and large-scale trace-driven evaluations conducted with thousands of Earth observation satellites and Starlink satellites that are currently in-orbit. The results of trace-driven evaluations confirm the superior performance of our design.
\end{itemize}

\section{BACKGROUND AND MOTIVATION}

\textbf{Earth Observation Satellites:} Modern Earth observation satellites working in constellations consist of hundreds of satellites \cite{LEONASA}. For instance, Planet’s Dove constellation for Earth imagery is composed of nearly 200 low-cost CubeSats (10$\times$10$\times$30 centimeters) with off-the-shelf components \cite{PlanetConstellation}. These small CubeSats typically orbit at altitudes of 370$\sim$430 km which is lower than communication consternation to capture terabytes of high resolution pictures per day\cite{planet_satellite_monitoring, 295557}. 
Current EO satellite data transfer largely follows a “wait and transfer” model, relying on a limited number of ground stations. Due to the high cost of building and maintaining ground stations, this approach often results in delays of several hours before collected data can be transmitted.

\textbf{LEO Satellite Constellation:} The emergence of large LEO constellations, comprising thousands of satellites, has been driven by lower launch and manufacturing costs of small satellites. 
LEO satellites are satellites that operate at altitudes typically ranging from about 160 km to 2,000 km above Earth's surface. Time to complete one orbit is approximately 90 to 120 minutes. For any ground observer, the satellite is visible for around ten minutes and has four to six good contacts every day. The widespread deployment of LEO satellite mega-constellations, such as Starlink \cite{Starlink} and Kuiper \cite{Kuiper}, has enabled global Internet coverage by ISL and high frequencies like the Ku or Ka band, with more than 5 million subscribers in over 100 countries as of Feb 2025 \cite{Wikipedia:Starlink}.

\begin{figure}
  \centering
  \includegraphics[width=0.55\linewidth]{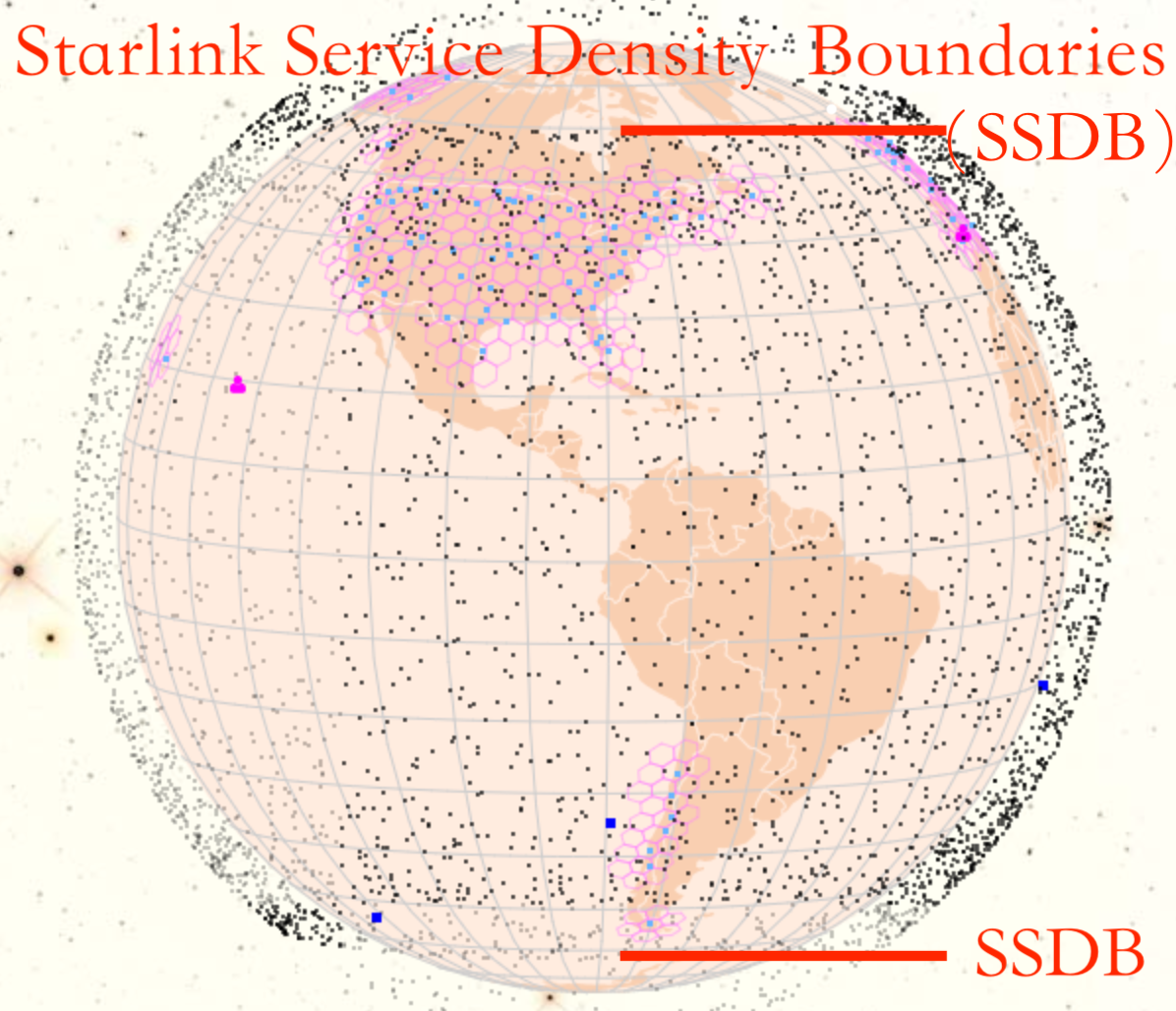}
  \vspace{-0.1in}
  \caption{Starlink LEO satellite constellation.}
  \label{fig:StarlinkSystem}
  \vspace{-0.3in}
\end{figure}

\textbf{Starlink:} As shown in Figure \ref{fig:StarlinkSystem}, Starlink is a LEO satellite constellation comprising over 7,000 satellites in orbit as of April 2025. Its architecture includes four key components: (1) in-orbit satellites, (2) user terminals, (3) ground stations, and (4) Points of Presence (PoPs). User terminals, installed on customer premises, connect to one satellite at a time, while each satellite can serve multiple terminals simultaneously. Ground stations, equipped with phased-array antennas, relay satellite traffic to PoPs via wired links.
Starlink satellites operate at altitudes ranging from 340 to 1,200 km, with most clustered around 550 km. Ground users typically connect to a fixed PoP due to their stationary location, resulting in a stable and predictable link. For satellite-to-satellite communication, Starlink employs both Ku/Ka band radio frequency (RF) and increasingly, since late 2023, laser-based inter-satellite links.


\textbf{Cost of Current EO data transfer:} Commercial EO data delivery remains costly due to the need for extensive ground infrastructure~\cite{geoawesome2023business}. Among leading providers such as Airbus \cite{airbus2025earthobservation}, Vantor \cite{vantor2025spatialintelligence}, and Planet Labs \cite{PlanetConstellation}, the latter is representative for its large-scale constellation of over 200 active satellites and broad global customer base~\cite{planetinvestor2025,igspace2023overview}. Planet relies on the \textit{wait-and-transfer} model using dedicated ground stations. According to its financial filings, ``Ground stations and ground station equipment'' are listed as fixed assets valued at \$21.27M (as of Jan.~31,~2025), depreciated over 3--10~years~\cite{planet10k2025}. In addition, annual service lease agreements of \$9.6M in FY2025 likely reflect maintenance costs. Based on an average daily downlink volume of 30~TB~\cite{planet_testimony2023,planet_leoconn2021}, the estimated unit cost ranges from \$1,524/TB over 3 years to \$1,071/TB over 10 years, placing traditional ground station delivery costs in the \$1,000–-\$1,500 per TB range.

\textbf{Ka-Band based Satellite Links:} 
Most commercial EO satellites use the X-band (8.025–8.400 GHz), designated by the International Telecommunication Union (ITU) for Earth Exploration Satellite Service (EESS), for data downlink due to its global regulatory protection and low atmospheric attenuation \cite{eo_xband_usage,ntrs2020_downlink}. However, The surge in EO data driven by high-resolution imagery, SAR products, and near real-time delivery, is straining the limited X-band spectrum.
Meanwhile, the Ka-band (25.5–27 GHz), though also allocated to EESS \cite{itu_sa1862,itu_handbook_spectrum}, remains under-utilized in commercial EO systems due to its heightened susceptibility to rain attenuation and technical complexity of Ka-band ground terminals \cite{nasa_smallsat_soa,ksat2018_whitepaper}. Fortunately, these limitations do not apply to Ka-band links between EO satellites and Starlink satellites.
Early Ka-band demonstrations (e.g., Planet’s 320 Mbps CubeSat \cite{planet_ka_cubesat,leveque2019ka}) and NASA’s NISAR mission (targeting 3.5 Gbps \cite{nasa_nisar_kaband}) highlight a shift toward higher-frequency solutions and the promising potential of Ka-band satellite communications.

The above aspects motivate exploring alternative EO data downlink architectures, such as leveraging Starlink’s  Ka-band infrastructure to alleviate spectrum bottlenecks and enable efficient, low-cost, and global EO data delivery. The rapid evolution of Starlink and emerging measurement studies now allow deeper investigation into system-level implementation challenges. 
While prior efforts like OrbitCast \cite{lai2021orbitcast} discussed high-level insights, our work provides a more detailed and systematic analysis. 
Though implemented on Starlink, our approach is broadly applicable to other modern LEO constellations.


\begin{figure*}[t]
  \centering

  \begin{subfigure}[t]{0.3\textwidth}
    \centering
    \includegraphics[width=\linewidth]{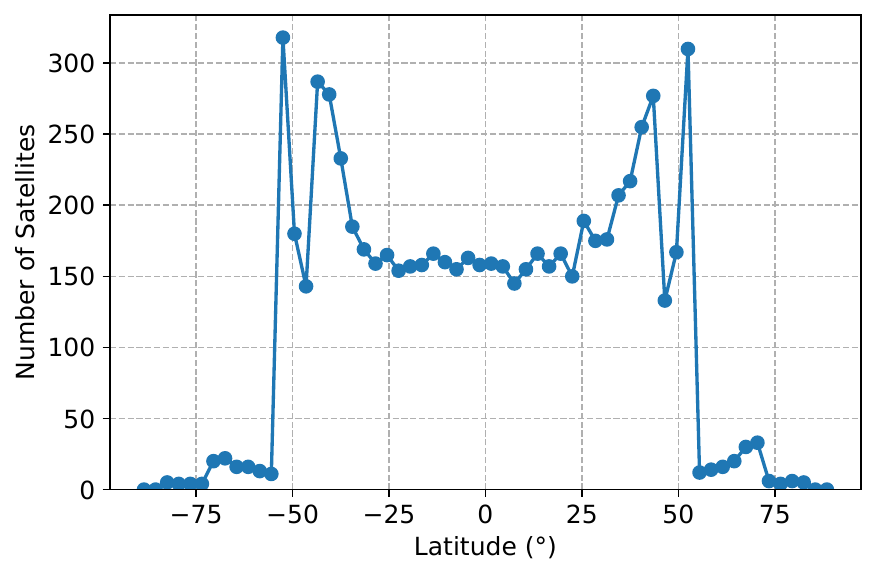}
    \vspace{-0.2in}
    \caption{Starlink satellite density.}
    \label{fig:StarlinkSatDensity}
  \end{subfigure}
  \hfill
  \begin{subfigure}[t]{0.3\textwidth}
    \centering
    \includegraphics[width=\linewidth]{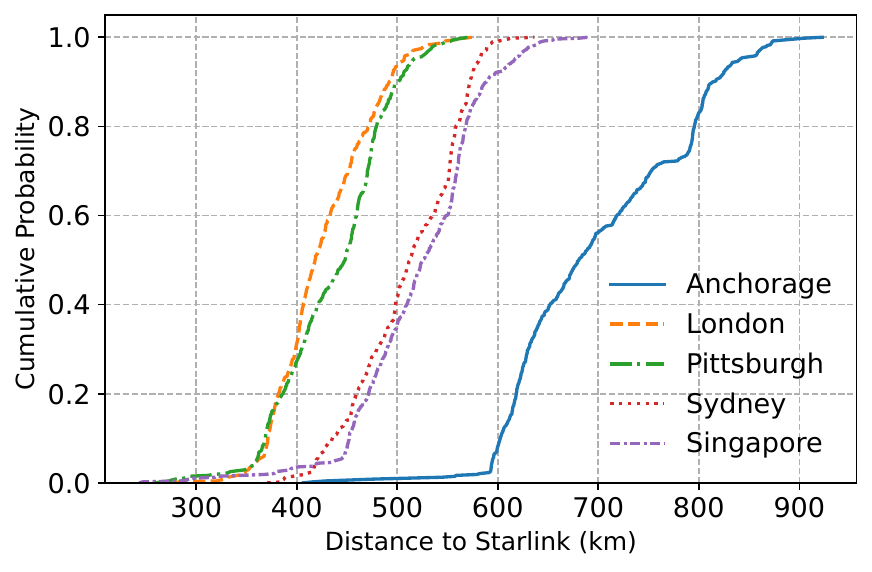}
    \vspace{-0.2in}
    \caption{CDF of ground-to-satellite distance.}
    \label{fig:GUdistance}
  \end{subfigure}
  \hfill
  \begin{subfigure}[t]{0.3\textwidth}
    \centering
    \includegraphics[width=\linewidth]{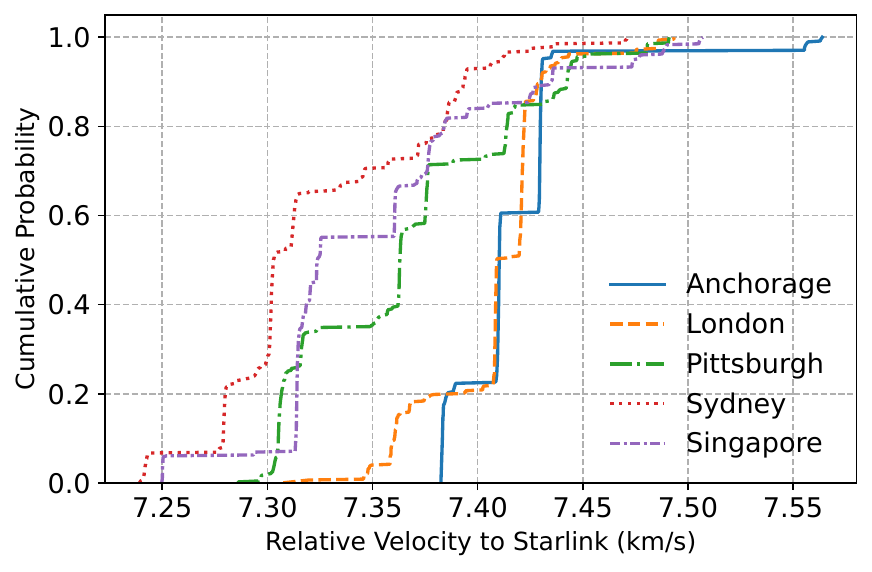}
    \vspace{-0.2in}
    \caption{CDF of relative velocity.}
    \label{fig:GUvelocity}
  \end{subfigure}
    \vspace{-0.15in}
  \caption{Starlink satellite distribution and CDFs of network parameters from 5 ground locations.}
  \vspace{-0.2in}
  \label{fig:ThreeCombinedPlots}
\end{figure*}

\section{PRELIMINARY ANALYSIS}
To study the feasibility of using Starlink satellites for EO data downlink, we conduct a preliminary analysis. We investigate the TLE (Two-Line Element) data \cite{space-track} for all 7,061 working Starlink satellites as of Mar. 1st, 2025 to simulate their orbit features. We also obtained TLE data for representative Planet EO satellites \cite{Planet} to track their position changes on the same day. To establish a stable connection to Starlink, three key factors are considered: (1) the satellites must be within communication range, (2) maintain a low relative velocity, and (3) exhibit minimal relative velocity variation over time \cite{alawieh2016inter}. 

\subsection{Starlink Global Distribution}
While Starlink provides global coverage, its satellite distribution is not geographically uniform. Using the TLE data of all active Starlink satellites, we compute the Starlink global satellite density distribution, and the result is plotted in Figure~\ref{fig:StarlinkSatDensity}. 
Two key patterns are observed. First, there are two distinct latitudinal boundaries that divide the globe into three zones: a central high-density region flanked by two low-density zones. This pattern arises because the Starlink network was initially optimized for regions within ±53° latitude, as also shown in Figure~\ref{fig:StarlinkSystem}. We refer to these latitudinal boundaries as the \textit{Starlink Service Density Boundaries (SSDB)}. 
Second, within the SSDB, satellite density generally decreases with decreasing latitude, except for a noticeable drop near the boundaries of the SSDB. The sparsest coverage is observed near the equator and around ±48°, indicating that the satellite distribution is not uniform even within the nominal high-density zone. The reduced density near the equator is primarily attributed to the orbital inclinations of the deployed satellites. The majority of Starlink satellites are placed in ~53° and ~43° inclined orbits, whose ground tracks intersect the equatorial region less frequently. As a result, satellite density gradually declines toward lower latitudes despite the network’s global coverage design. The observed sparsity around ±48° is due to this region lying between the coverage envelopes of the ~53° and ~43° orbital shells, where only satellites from the ~53° inclination are present, and those from the ~43° inclination have not yet reached. 

\subsection{Ground Users and Starlink Satellites}

To evaluate the link capability of the Starlink network, we use ground users as reference points. Five representative locations across different latitudes are selected, i.e., Pittsburgh, Singapore, London, Sydney, and Anchorage. We compute the distance and relative velocity between each location and its nearest Starlink satellite over a two-hour period (covering more than one orbital cycle of a typical LEO satellite). This enables an assessment of the temporal stability and global consistency of Starlink coverage.

The results are reported in Figure~\ref{fig:GUdistance} and Figure~\ref{fig:GUvelocity}. We find that the relative velocity between ground users and the nearest Starlink satellite remains stable, around 7.4~km/s, which 
we refer to as \textit{Starlink Ground Link Velocity} ($v_G$=7.4 km/s), representing the typical relative velocity under which stable communication is maintained.
Slightly larger variations are observed in Singapore and Anchorage. In contrast, the distance to the nearest satellite varies with latitude: Pittsburgh and London experience ranges of 350$\sim$550~km; Sydney, 400$\sim$600~km; Singapore, 400$\sim$700~km; and Anchorage, 600$\sim$900~km. 
This observation is consistent with the known distribution of the Starlink constellation, which leads to sparser satellite coverage near the equator. As a result, users located at low latitudes have fewer nearby satellites available at any given time. Beyond the SSDB, satellite availability decreases even more significantly. For example, Anchorage, located outside the SSDB, experiences comparatively longer distances to the nearest Starlink satellites, which is expected given the design characteristics of the network.
Despite experiencing the largest distance range, Anchorage in practice still maintains stable Starlink connectivity. This indicates that reliable communication is achievable at distances up to 900~km for ground users, provided the relative velocity remains close to 7.4~km/s with limited fluctuation.

\subsection{Space Users and Starlink Satellites}
Given the Starlink performance we learnt from ground users, can they directly support space users, e.g., EO satellites, without upgrading the network and communication capacity?
Using the Planet EO satellites \cite{Planet} as examples, we analyze their distance and relative velocity to the 7,061 working Starlink satellites over multiple orbital periods. 

\begin{figure}[t]
  \centering
  \begin{subfigure}{0.48\columnwidth}
    \includegraphics[width=\linewidth]{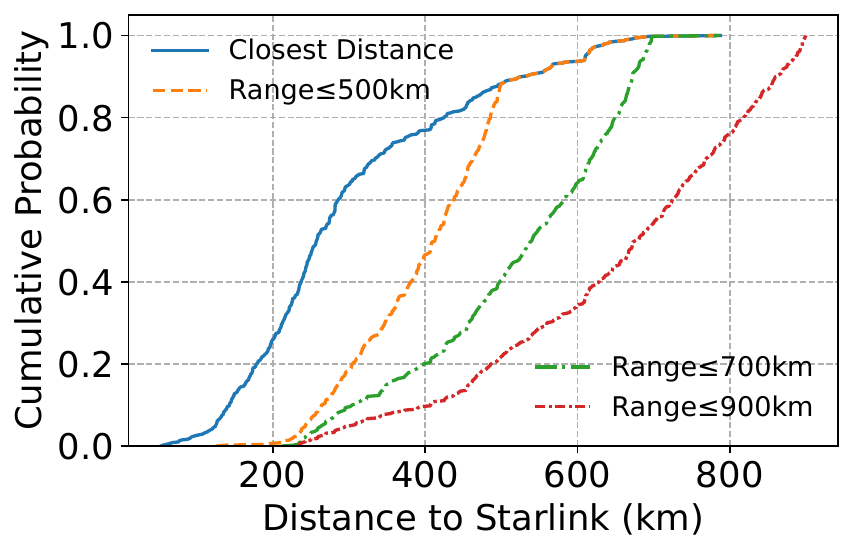}
    \vspace{-0.2in}
    \caption{CDF of distance.}
  \end{subfigure}
  \hfill
  \begin{subfigure}{0.48\columnwidth}
    \includegraphics[width=\linewidth]{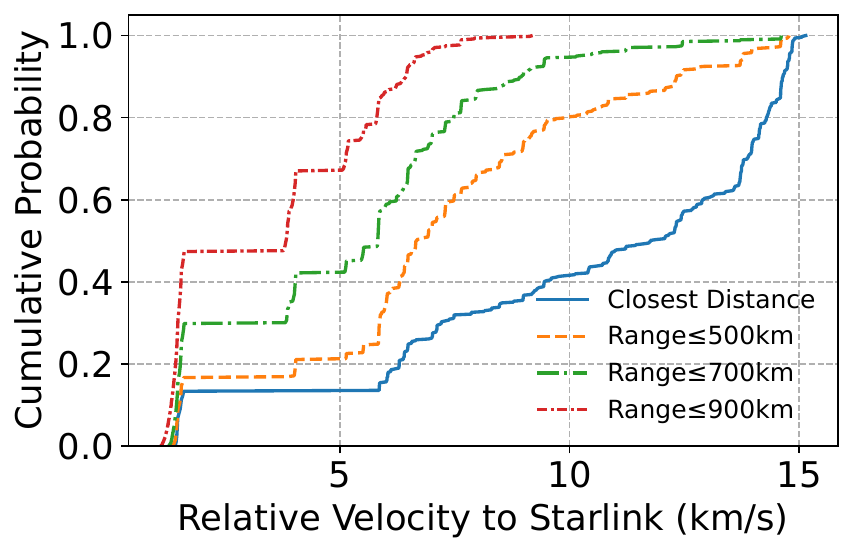}
    \vspace{-0.2in}
    \caption{CDF of relative velocity.}
  \end{subfigure}
  \vspace{-0.12in}
  \caption{CDF of distance and relative velocity from EO satellites to Starlink for different selection criteria.}
  \label{fig:su_strategy_cdf}
  \vspace{-0.3in}
\end{figure}

\begin{figure}
  \centering
  \includegraphics[width=0.7\linewidth]{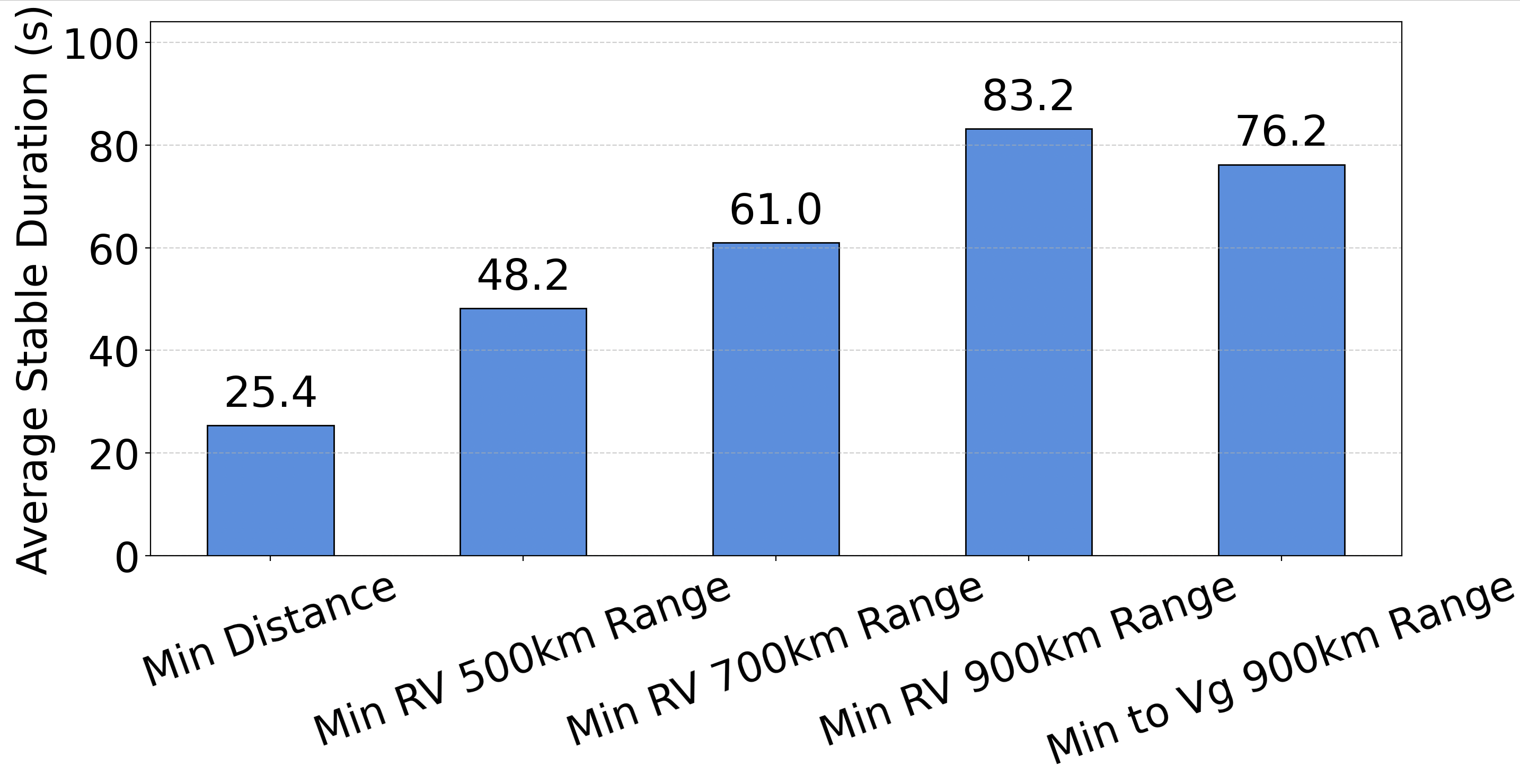}
  \vspace{-0.15in}
  \caption{Average connection time per Starlink satellite for different criteria (RV: relative velocity).}
  \vspace{-0.15in}
  \label{fig:AverageLinkTime}
\end{figure}

We first use two different selection criteria to search the nearby Starlink satellite for connection, i.e., (1) the nearest Starlink satellite, and (2) the Starlink satellite with minimal relative velocity within 500 km, 700 km, and 900 km to the EO satellite, respectively.
Figure~\ref{fig:su_strategy_cdf} reports the distance and relative velocity between the EO satellite and the chosen Starlink satellite under these searching criteria.
Figure \ref{fig:AverageLinkTime} shows the average connection time per each Starlink satellite before handover.
When always selecting the nearest Starlink satellite, the distance remains below 400~km for 80\% of the time. However, the relative velocity can reach up to 14.5~km/s, and the average connection time per Starlink satellite is 25.4 seconds.
When we gradually expand the search radius to 500 km, 700 km, and 900~km, and selecting the satellite with the lowest relative velocity within that range, both the highest and average relative velocity decrease. The average connection time per satellite is much longer than connecting to the nearest one, indicating lower operation cost. 
However, the variation in the relative velocity is still considerable. 


\begin{figure}[t]
  \centering
  \begin{subfigure}{0.48\columnwidth}
    \includegraphics[width=\linewidth]{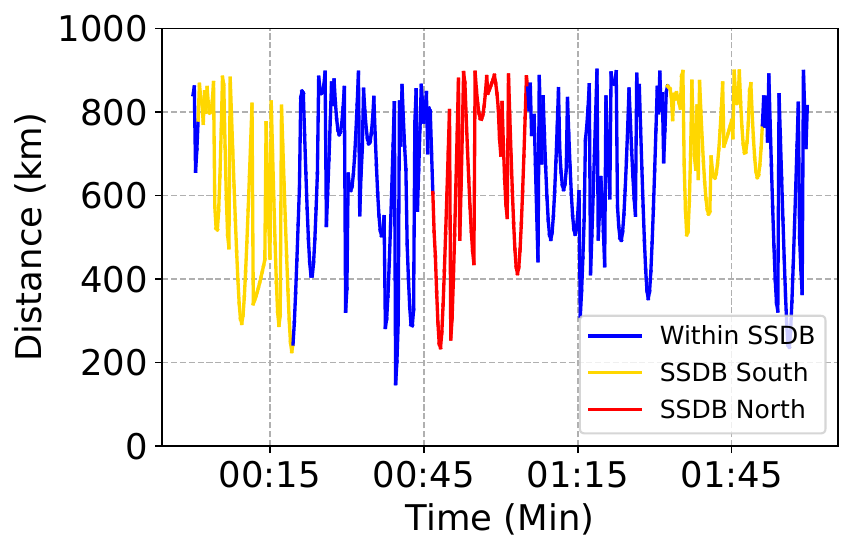}
    \vspace{-0.2in}
    \caption{Distance to Starlink.}
  \end{subfigure}
  \hfill
  \begin{subfigure}{0.48\columnwidth}
    \includegraphics[width=\linewidth]{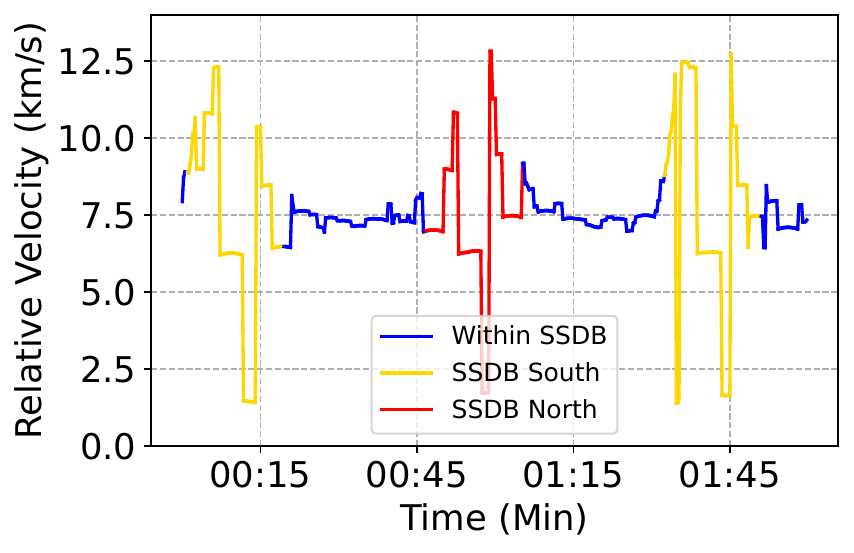}
    \vspace{-0.2in}
    \caption{Relative velocity.}
  \end{subfigure}
  \vspace{-0.12in}
  \caption{Distance and relative velocity from EO satellite to Starlink satellites with $min|v-v_G|$ in 900 km.}
  \vspace{-0.2in}
  \label{fig:su_SGLV_cdf}
\end{figure}

Inspired by the $v_G$=7.4 km/s as a globally used relative velocity to maintain stable connection for ground users, we adapt the selection criterion to find the Starlink satellite whose relative velocity is closest to the $v_G$ within the specific search range. Figure~\ref{fig:su_SGLV_cdf} plots the distance and relative velocity $v$ from a typical EO satellite to the Starlink satellite selected via the adapted search criterion, i.e., the one with $min|v-v_G|$ within 900 km.
When the EO satellite is located within the SSDB, we find that Starlink satellites with relative velocities close to the $v_G$ can be reliably found. 
However, as the EO satellite approaches or crosses the SSDB, the relative velocity becomes more irregular and varies significantly, and the $v_G$-based criterion performs poorly.
We then dig into the regions where EO satellites are close to the poles (outside the SSDB). Interestingly, in such scenarios, when selecting the Starlink satellite with the lowest relative velocity (within 900~km), we observe that it maintains a consistently low relative velocity, around 1.5~km/s, for most of the cases. 
We attribute the consistently low relative velocity between Starlink and EO satellites outside the SSDB to the similarity in their orbital planes, particularly in polar or near-polar configurations. This orbital alignment reduces the relative motion between the satellites, thereby enabling stable communication opportunities.

\begin{figure}[t]
  \centering
  \begin{subfigure}{0.48\columnwidth}
    \includegraphics[width=\linewidth]{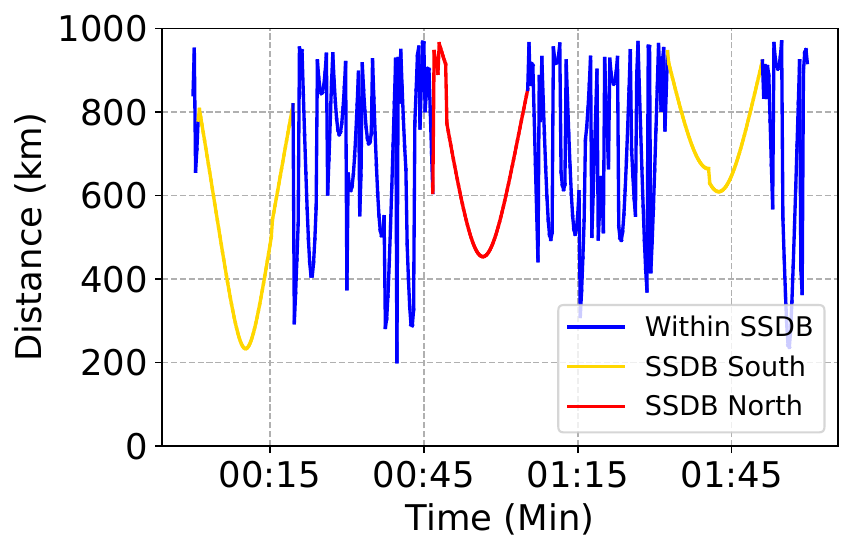}
    \vspace{-0.2in}
    \caption{Distance to Starlink.}
  \end{subfigure}
  \hfill
  \begin{subfigure}{0.48\columnwidth}
    \includegraphics[width=\linewidth]{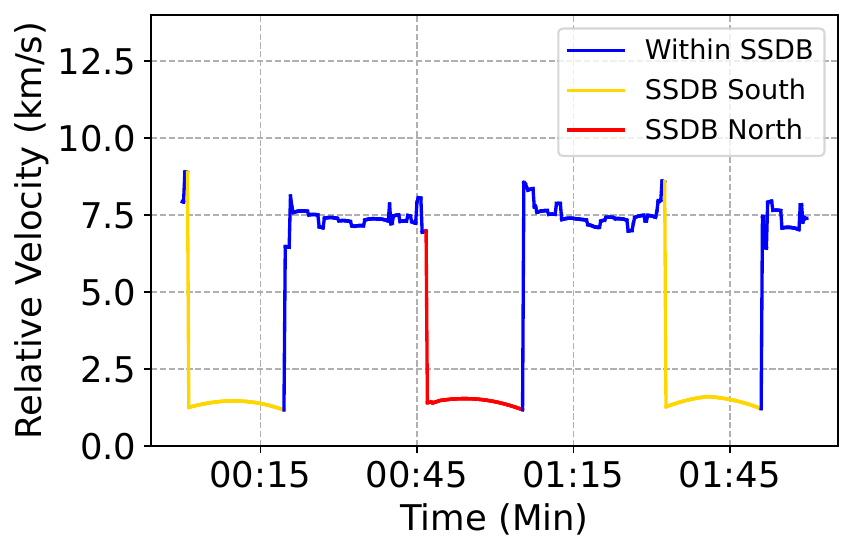}
    \vspace{-0.2in}
    \caption{Relative velocity.}
  \end{subfigure}
  \vspace{-0.12in}
  \caption{Distance and relative velocity from EO satellite to Starlink satellites with dual-criterion selection.}
  \vspace{-0.2in}
  \label{fig:DualCriGood_cdf}
\end{figure}

Motivated by this observation, we may further upgrade the selection criterion and implement different search criteria when EO satellites are inside and outside the SSDB, respectively.
Specifically, we can select the Starlink satellite with $min|v-v_g|$ when the EO satellite is within the SSDB, and the one with lowest relative velocity when the EO satellite is outside the SSDB.
Meanwhile, considering space users build links in space and has no atmosphere attenuation, we expand the search range to 968~km to account for estimated atmospheric losses under typical cloudy weather conditions at a 45° elevation angle. This is reasonable, as Starlink selects satellites with angles of elevation (AOEs) between 45° and 90° for over 80\% of the time \cite{tanveer2023makingsenseconstellationsmethodologies}. 
Figure~\ref{fig:DualCriGood_cdf}(b) illustrates the resulting velocity distribution under this dual-criterion strategy in the range of 968 km. We see that the relative velocity is consistently stable across the three regions.
The average connection time per Starlink satellite (Figure \ref{fig:AverageLinkTime}) is also comparable to other strategies, if not better.
The results demonstrate that by applying different selection criteria outside and inside the SSDB, it is possible to maintain stable relative velocities within a feasible range for communication via the Starlink constellation.

\begin{figure}[t]
  \centering
  \begin{subfigure}{0.49\columnwidth}
    \includegraphics[width=\linewidth]{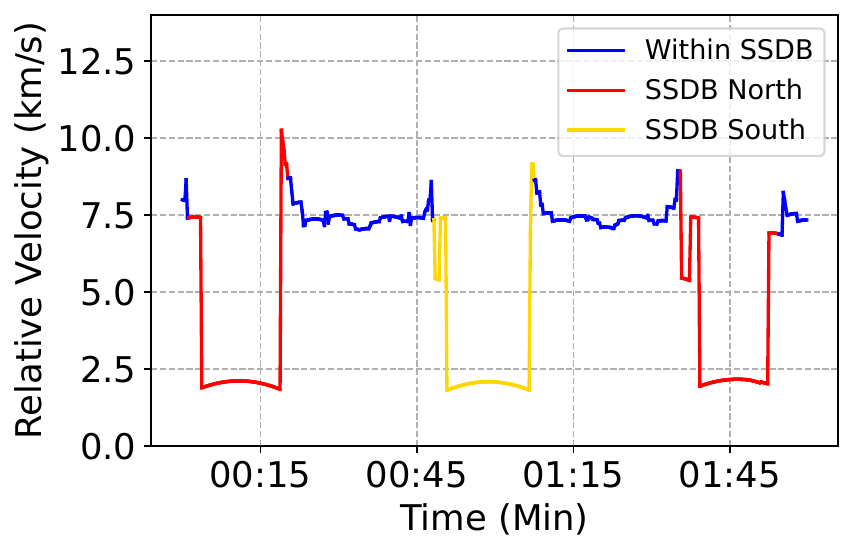}
    \vspace{-0.2in}
    \caption{Planet FLOCK 4Q1.}
  \end{subfigure}
  \hfill
  \begin{subfigure}{0.49\columnwidth}
    \includegraphics[width=\linewidth]{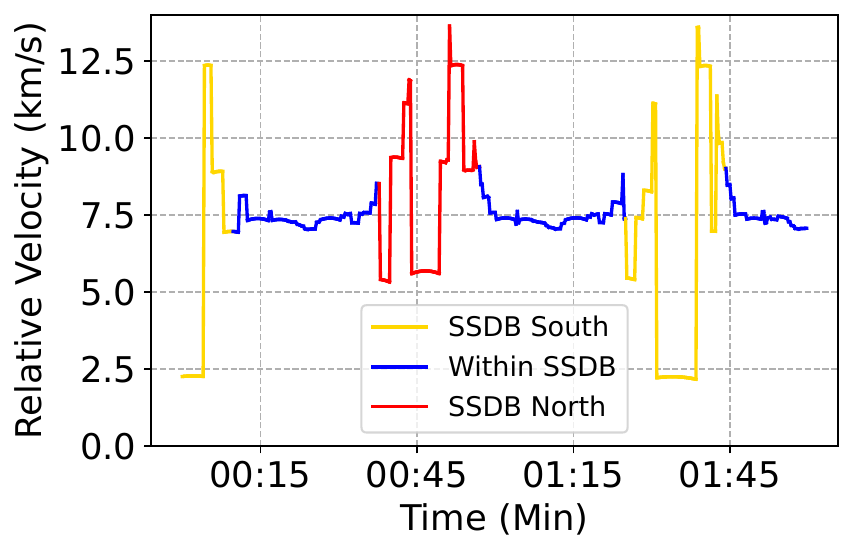}
    \vspace{-0.2in}
    \caption{Planet FLOCK BE20.}
  \end{subfigure}
  \vspace{-0.15in}
  \caption{Examples of two Planet satellites with relative velocity fluctuations with dual-criterion selection.}
  \label{fig:DualCriBad_cdf}
  \vspace{-0.08in}
\end{figure}

We also observe that not every EO satellite outside the SSDB can always find a Starlink satellite with a sufficiently aligned orbit. The Planet Flock 4Q-1 and Flock 4BE-20 are two examples. The relative velocity between them and Starlink is depicted in Figure~\ref{fig:DualCriBad_cdf}(a) and Figure~\ref{fig:DualCriBad_cdf}(b), respectively. 
We see there are some fluctuations in the relative velocity when Flock 4Q-1 is crossing the SSDB. 
For the Flock 4BE-20 satellite, the variation of relative velocity is significantly high when it is outside the SSDB.
However, such cases account for only around 10\% of all EO satellites we investigated. 

\section{STARLINK AIDED EARTH OBSERVATION DATA TRANSFER}

\begin{figure}
  \centering
  \includegraphics[width=0.95\linewidth]{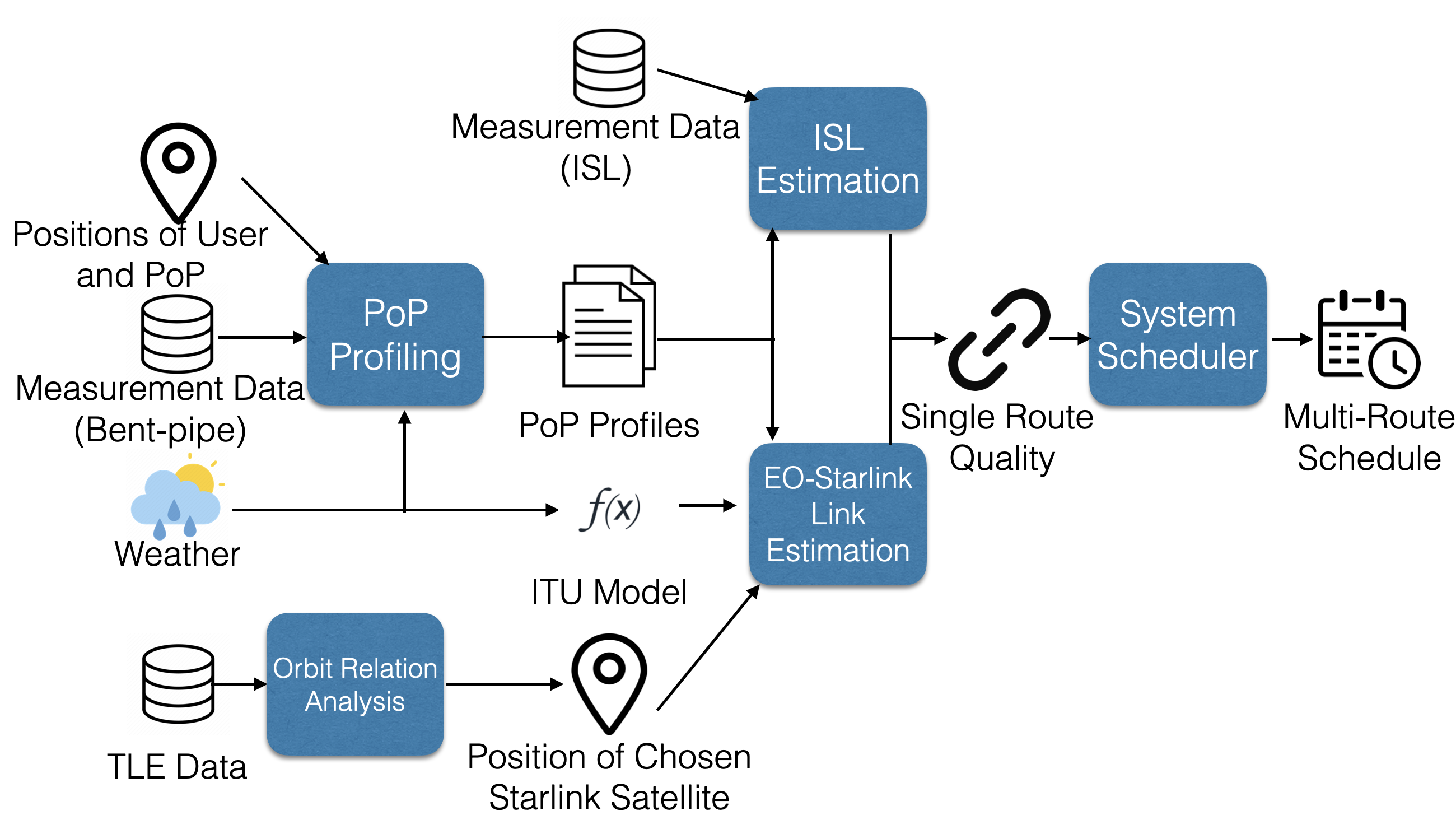}
  \vspace{-0.1in}
  \caption{SSU System architecture.}
  \label{fig:SystemOverview}
\end{figure}

Based on preliminary analysis, we propose a Starlink-aided Earth observation data transfer system (Figure~\ref{fig:ssu}), where EO satellites act as Starlink space users, dynamically assigned to Starlink satellites to relay data to PoPs via ground stations. The SSU system architecture (Figure~\ref{fig:SystemOverview}) includes several key modules: the PoP profiling module (Sec. 4.1) uses machine learning to predict PoP performance across locations and weather conditions; the ISL estimation module provides baseline metrics for inter-satellite link loss and delay; the Orbit Relation Analysis module (Sec. 4.2) evaluates link quality based on distance and relative velocity between EO and Starlink satellites; and the EO-Starlink link estimation module (Sec. 5.2) assesses link quality between EO satellites and selected Starlink nodes. These modules feed into the system scheduler (Sec. 4.3), which plans connections and handovers to optimize throughput. The resulting schedule is transmitted through the Starlink network to the relevant satellites, and EO satellites follow this schedule during orbit to transmit data to the designated PoPs based on their location.

\subsection{Profiled PoP Selection}
Starlink ground users typically connect to a fixed PoP, but in SSU, the constantly shifting spatial relationship and routing between moving EO satellites and PoPs requires dynamic adaptation. Due to Starlink’s proprietary nature and lack of real-time in-network access, we develop a prediction framework to select suitable PoPs without relying on in-network measurements.

\setlength{\textfloatsep}{5pt}

\subsubsection{Route Quality Analysis}
The loss rate and delay for a packet transferring from an EO satellite to a PoP includes: 
(1) Propagation loss between observation satellite and its served Starlink satellite. It's nearly proportional to the distance between the two satellites in space. 
(2) Starlink in-network delay and loss, stemming from the ISL transport, propagation from the last Starlink satellite to the ground station (GS) (affected by the weather resulting in 0$\sim$10 dB Signal to Noise Ratio variation \cite{shrestha2017characterization}), and GS-to-PoP transmission. 
(3) Hardware effect including the processing of satellites hardware, the design of the antennas and circuits in the transmitter and receiver.
While propagation loss and ISL delay can be estimated using orbital models and light-speed assumptions \cite{ITU1, ITU2, ITU3}, overall delay depends on factors like ISL hop count, and the exact routing within Starlink remains a black box.
 

\subsubsection{PoP Profiling and ISL Estimation}
We use (1) the large Starlink measurement dataset LENS \cite{zhao2024lens}, (2) locations of ground users and destination PoPs, and (3) local weather conditions \cite{Weather-forecasting} to train models that characterize communication performance. PoP profiling focuses on users near a PoP using bent-pipe routing (user to satellite, down to a ground station, then to the PoP), allowing measurements without Starlink’s internal access. The goal is to predict packet loss and delay based on location and weather. ISL estimation examines how inter-satellite links affect routing quality for users far from a PoP, analyzing scenarios where users switch PoPs and studying the resulting quality changes.


\textbf{Model architecture:} 
SSU uses an ensemble model to predict route quality by combining gradient-boosted regression trees and deep learning-based regression for complementary insights. The gradient-boosted component employs XGBoost \cite{Chen_2016} with 1000 trees (max depth 15) trained on five folds of the feature matrix, while the deep model has six ReLU-activated layers for non-linearity. Outputs from both models are merged and passed through a final linear regression layer to estimate packet loss rate and delay.


\textbf{Input features:} We select features that capture key factors affecting signal quality, including positional attributes (latitude and longitude of users and ground stations) and detailed weather data such as precipitation, temperature, pressure, humidity, cloud cover, wind speed, and snow. Service-specific attributes like user tier, dish revision, and obstruction ratio are also incorporated. These combined inputs enable the model to predict packet loss and delay, and estimate data transmission rates using known Starlink bandwidth settings.

\textbf{Loss function and methodology:} The models are trained using the mean squared error (MSE) loss function. 
First, for predicting packet loss and delay which are critical for SSU’s cloud upload use case, we focus on the uplink from ground user to PoP, calculating loss rates every second to align with Starlink’s 15-second handover scheme \cite{tanveer2023makingsenseconstellationsmethodologies}. 
Second, ISL estimation involves three steps: (1) measuring quality differences before and after PoP switching, (2) isolating PoP-specific effects using profiles of the two PoPs, and (3) subtracting these effects to extract the impact of ISL distance on route quality. We assume ISL impact is proportional to transmission distance under ideal space conditions.

In short, we profile three key factors for selecting the optimal PoP for a space user:
(1) the PoP’s identity,
(2) local weather conditions at the PoP, and
(3) the distance to the PoP, which influences ISL-related loss and delay.


\subsection{Orbit Analysis Based Link Selection}
As shown in our preliminary analysis, space users often have shorter communication distances than ground users but much higher relative velocities. Starlink employs Orthogonal Frequency Division Multiplexing (OFDM) \cite{humphreys2023signal}, and the high Doppler shifts induced by this relative motion can lead to significant frequency deviations, resulting in adjacent channel interference.
Even with our dual-criterion strategy, there are instances where the relative velocity can exceed the $v_G$ to around 8.5 km/s, as illustrated in Figure~\ref{fig:DualCriGood_cdf}. 
 To address this, we design an orbit-aware link selection strategy that chooses the optimal satellite by quantifying the combined impact of frequency shift and distance.

For the distance, we adopt the Free-Space Path Loss model:
\begin{equation}
\text{FSPL (dB)} = 20 \log_{10}(d) + 20 \log_{10}(f) + 20 \log_{10}\left( \frac{4\pi}{c} \right),
\end{equation}
where \( d \) is the distance between the transmitter and receiver, \( f \) is the carrier frequency, and \( c \) is the speed of light in vacuum.
To address signal degradation from frequency shifts, we analyze SNR loss within the OFDM framework using the model in \cite{moose1994technique}, which characterizes the impact of Carrier Frequency Offset (CFO) on OFDM systems.
The SNR degradation caused by CFO is given by:
\begin{equation}
\text{SNR} \geq \frac{\frac{E_c}{N_0} \left( \frac{\sin(\pi \varepsilon)}{\pi \varepsilon} \right)^2}
{1 + 0.5947 \left( \frac{E_c}{N_0} \right) \left( \sin(\pi \varepsilon) \right)^2},
\quad |\varepsilon| < 0.5,
\end{equation}
where \( \frac{E_c}{N_0} \) denotes the ideal energy-to-noise ratio that a subcarrier can achieve in the absence of frequency offset, and \( \varepsilon \) is the normalized CFO, defined as the frequency shift divided by the subcarrier symbol rate. When \( |\varepsilon| > 0.5 \), the subcarrier shifts into adjacent channels, resulting in severe inter-carrier interference, rendering traditional approximations ineffective and potentially making the link unusable.

The Starlink system tolerates CFO up to a threshold corresponding to the \textit{maximum relative velocity observed for ground users}, denoted as \( v_{\text{G-max}} \). The velocity that can make the adjacent channel interference is \( v_{\text{critical}} \). Our Starlink satellite link selection strategy is shown in Algorithm~\ref{alg:sat-selection}.

\begin{algorithm}[t!]
\caption{Satellite Selection Algorithm}
\label{alg:sat-selection}

\KwIn{Set of candidate satellites $S$, velocity thresholds $v_{\text{G-max}}, v_{\text{critical}}$}
\KwOut{Selected optimal satellite $s^*$}

\ForEach{satellite $s \in S$}{
    Compute distance $d_s$ and relative velocity $v_s$\;
}

\eIf(\tcp*[f]{Minimize path loss}){$\exists s \in S : v_s \leq v_{\text{G-max}}$}{
    $s^* \leftarrow \arg\min_{s}\{d_s \mid v_s \leq v_{\text{G-max}}\}$\;
}{
    \eIf(\tcp*[f]{Prioritize velocity}){$\forall s \in S: v_s > v_{\text{critical}}$}{
        $s^* \leftarrow \arg\min_{s}\{v_s\}$\;
    }(\tcp*[f]{Balance velocity and distance (maximize SNR)}){
        $s^* \leftarrow \arg\max_{s}\{\text{SNR}(v_s, d_s)\}$ using Eq.~(2)\;
    }
}
\Return $s^*$\;
\end{algorithm}

In the link selection process, we apply the dual-criterion strategy on top of this general orbit-aware link selection framework.
When the space user is inside the SSDB, it connects to Starlink satellites with relative velocity near $v_G$. Since these candidate satellites may vary in both distance and velocity and the velocity can sometimes exceed \( v_{\text{G-max}} \), they are ranked based on our link selection strategy. 
A handover may occur either due to rank changes or as part of system-level optimization (See 4.3).
Outside the SSDB, the selection prioritizes satellites with minimal relative velocity, requiring Doppler compensation to shift from $v_G$ (7.4 km/s) to a much lower velocity (approximately 1.4 km/s). Among the candidates with minimal relative velocity, those with the shortest communication distance are preferred to maximize the SNR.
If no Starlink satellite with a sufficiently similar orbital plane is available, the space user is considered temporarily disconnected.

\subsection{System Scheduler}

Building on Sections 4.1 and 4.2, estimating the quality of a single route is straightforward once a Starlink satellite and destination PoP are assigned. However, link performance varies with orbital changes, and resource allocation must address competition for Starlink access. Key factors include: (a) SSU’s role as a global space agency integrated with Starlink to manage links for multiple providers, (b) rapid growth in observation satellites increasing demand for limited resources, and (c) Starlink’s expanding ground user base competing for the same links, with frequent handovers every 15 seconds \cite{tanveer2023makingsenseconstellationsmethodologies}. Thus, scheduling becomes a dynamic allocation of link matrices between observation and Starlink satellites for each stable session under highly competitive conditions to maximize overall throughput.

\setlength{\tabcolsep}{3pt}  
\begin{table*}[h!]
    \centering
    \small
    \begin{tabular}{|l|l|l|c|l|l|}
        \hline
        \textbf{Dish ID} & \textbf{Location} & \textbf{Hardware Version} & \textbf{Sky Obs Ratio (\%)} & \textbf{PoP} & \textbf{Service Tier} \\ \hline
        \textit{victoria\_active\_1} & Victoria, BC, CA & \textit{rev3\_proto2} & 0.264 & Seattle & Standard \\ \hline
        \textit{victoria\_active\_2} & Victoria, BC, CA & \textit{rev3\_proto2} & 0 & Seattle & Mobile \\ \hline
        \textit{victoria\_inactive} & Victoria, BC, CA & \textit{rev3\_proto2} & 0 & Seattle & Inactive Mobile, Roam \\ \hline
        \textit{vancouver} & Vancouver, BC, CA & \textit{rev2\_proto3} & 4.564 & Seattle & Standard \\ \hline
        \textit{seattle} & Seattle, WA, USA & \textit{rev3\_proto2} & 10.198 & Seattle & Standard \\ \hline
        \textit{seattle\_hp} & Seattle, WA, USA & \textit{hp1\_proto1} & 0.257 & Seattle & Priority \\ \hline
        \textit{ottawa} & Ottawa, ON, CA & \textit{rev3\_proto2} & 13.961 & New York & Standard \\ \hline
        \textit{iowa} & Iowa City, IA, USA & \textit{rev1\_pre\_production} & 0.516 & Chicago & Standard \\ \hline
        \textit{denver} & Denver, CO, USA & \textit{rev3\_proto2} & 0.071 & Denver & Mobile, Roam \\ \hline
        \textit{louvain} & Louvain, Belgium & \textit{rev3\_proto2} & 0.027 & Frankfurt & Standard \\ \hline
        \textit{seychelles} & Seychelles & \textit{rev3\_proto2} & 0.646 & Lagos / Frankfurt & Mobile, Roam \\ \hline
        \textit{alaska} & Anchorage, AK, USA & \textit{rev3\_proto2} & 0.029 & Seattle & Mobile \\ \hline
        \textit{dallas} & Oxford, MS, USA & \textit{rev3\_proto2} & 15.979 & Dallas & Inactive Standard \\ \hline
    \end{tabular}
    \caption{Starlink dishes and locations used for profiling and validation.}
        \vspace{-0.3in}
    \label{tab:starlink_dishes}
\end{table*}

Drawing parallels between this problem and circuit scheduling in data centers \cite{10.1145/3230718.3230722, Celik2016-kz}, as well as hybrid ground station scheduling \cite{L2D2}, we develop a greedy algorithm tailored to optimize system-wide throughput.
Consider a set of observation satellites which are denoted as $\mathcal{O} = \{o_1, o_2, \ldots, o_M\}$ and a set of Starlink satellites $\mathcal{S} = \{s_1, s_2, \ldots, s_N\}$. The objective is to determine a sequence of linking matrices $P^t$ for $t = 0, \ldots, T$, where $T = 15$ seconds represents the minimum stable session duration for the system. Each matrix $P^t$ is an $N \times M$ matrix, where the entry in row $y$ ($y \leq N$) and column $x$ ($x \leq M$) indicates the link state between the $x$-th observation satellite and the $y$-th Starlink satellite. If the satellites are linked, the value is 1; otherwise, it is 0. 
It is important to note that each Starlink satellite can serve multiple users simultaneously, although SpaceX has not disclosed the exact maximum number of connections per satellite. As a result, multiple entries within a single row can be equal to 1. To calculate the current load on each Starlink satellite, we define an all-ones matrix $J^t_{M \times 1}$ and compute the link control matrix $C^t$ as $C^t = P^t \times J^t$. Each element in $C^t$ represents the number of active links on the corresponding Starlink satellite.

In the event of congestion, a constraint matrix $L^t$ is introduced, which contains $N$ values representing the remaining number of available links for each Starlink satellite. To ensure that the link constraints are respected, an element-wise comparison between matrices $C^t$ and $L^t$ is performed, enforcing the condition $C^t \leq L^t$.
For the $i$-th observation satellite with a backlog of $X_i^t$, let $D^t_{ij}$ represent the maximum data that can be transmitted between observation satellite $o_i$ and Starlink satellite $s_j$ at time $t$, given the current service tier. We then define the value matrix $\Phi^t$, where each element $\Phi^t_{ij} = \phi(\min(X_i^t, D^t_{ij}), t)$. If the observation satellite $o_i$ has sufficient data to transmit, it will transmit $D^t_{ij}$; otherwise, it transmits all available data $X_i^t$. Thus, $\Phi^t_{ij}$ reflects the actual value of transmitting data during session $t$ between observation satellite $o_i$ and Starlink satellite $s_j$.

The goal of the problem is to identify a sequence of matching matrices $P^t$ that maximizes the total value over time, subject to the constraint $C^t \leq L^t$:

\[
\max_{P^0, P^1, \ldots, P^T} \sum_t \text{Tr}(P^t \Phi^t),
\]
where $\text{Tr}(\cdot)$ denotes the sum of the main diagonal elements of a square matrix.
To address the throughput loss caused by switching delays, we define a penalty matrix $B^t$, where $B^t_{ij} = b \times \Phi^t_{ij}$. Here, $b \in \mathbb{R}^+$ represents the proportion of the switching delay relative to the stable session interval (15 seconds). The algorithm aims to select $P^t$ such that:
\[
P^t = \arg \max_{P^t} \text{Tr}(P^t (\Phi^t - B^t)).
\]
This optimization problem can be mapped to a maximum bipartite graph matching problem and solved using the Hungarian algorithm \cite{kuhn1955hungarian, munkres1957algorithms}.

\section{EXPERIMENTS}


\subsection{Route Quality Measurements}


\begin{figure}
  \centering
  \includegraphics[width=0.8\linewidth]{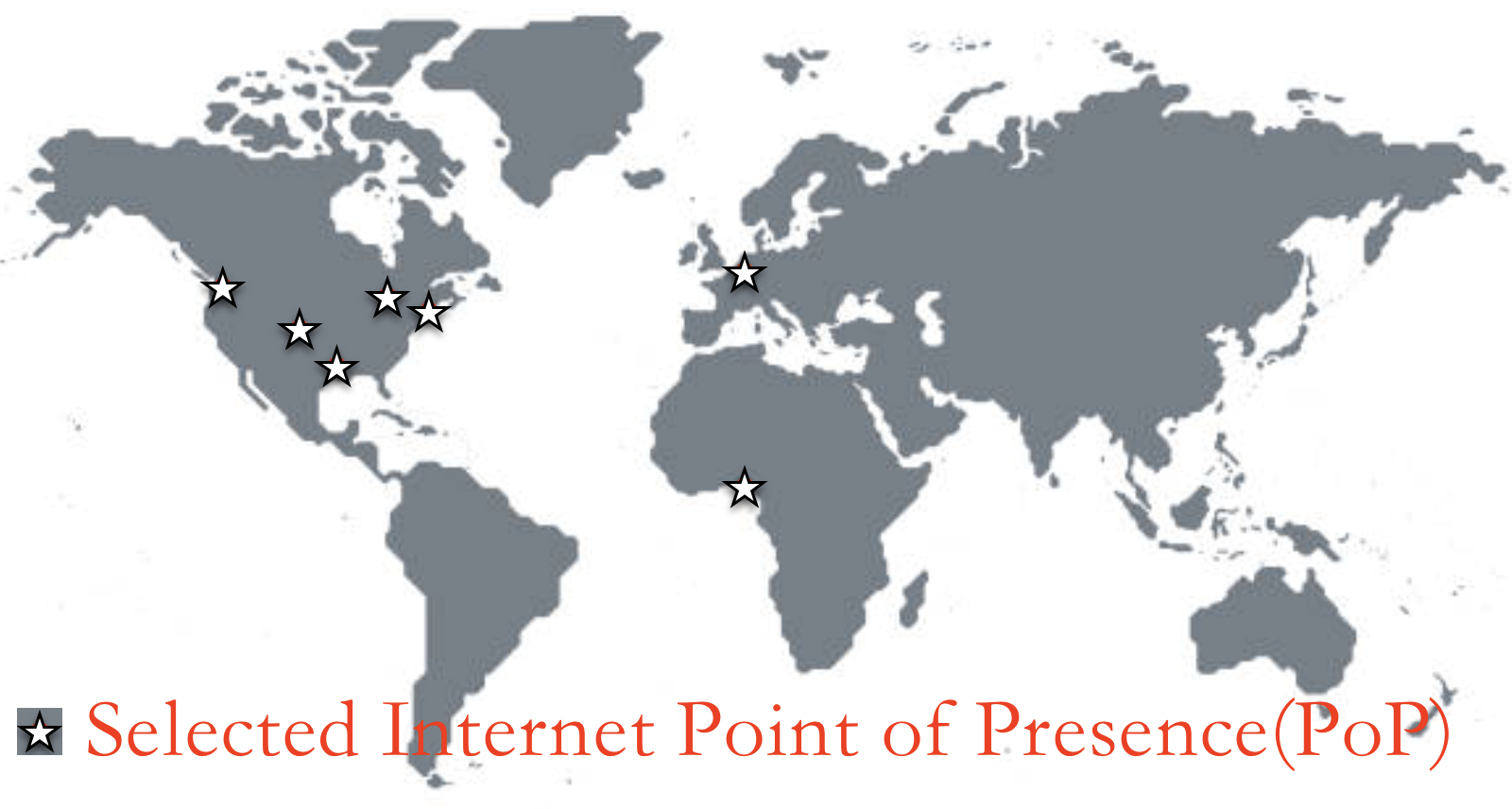}
  \vspace{-0.1in}
  \caption{Selected Starlink PoP distribution.}
  \label{fig:PoPWorld}
\end{figure}

To evaluate SSU’s ability to predict the quality of individual routes, we use real-world data from 13 Starlink dishes worldwide, associated with 7 PoPs across 3 continents: Africa, North America, and Europe (see Table~\ref{tab:starlink_dishes} \cite{zhao2024lens} and Figure \ref{fig:PoPWorld}).
We augmented these measurement data by integrating weather information obtained from the OpenWeather API \cite{OpenWeather}, which includes humidity, cloud coverage, precipitation (rain and snow), wind speed, temperature, and atmospheric pressure.

Positioned in the Western Indian Ocean without any ground stations within 5,000 km, the Starlink dish in Seychelles relies exclusively on ISLs for connectivity. Initially linked to the Lagos PoP in Nigeria, its association was switched to the Frankfurt PoP on Dec. 08, 2023. To estimate the impact of ISL on communication quality, we first calculate and record the packet loss rate to Lagos from this ground user located in Seychelles for 23 days. 
Next, to isolate the specific impact of ISL, we replicate the weather and dish conditions of the Seychelles user in the profiling of a terrestrial user near the PoP in Frankfurt. This setup allows us to estimate the hypothetical loss rate if the Seychelles user were situated close to the Frankfurt PoP and thus not reliant on ISL for data transmission. This mean of estimated loss rate is 0.27\%, providing an understanding of the loss rate attributed to local facilities and environmental conditions around the Frankfurt PoP.


\subsection{Space User to Starlink Link Estimation}

\begin{wrapfigure}{R}{0.25\textwidth}
  \begin{center}
    \vspace{-0.20in}
    \includegraphics[width=0.24\textwidth]{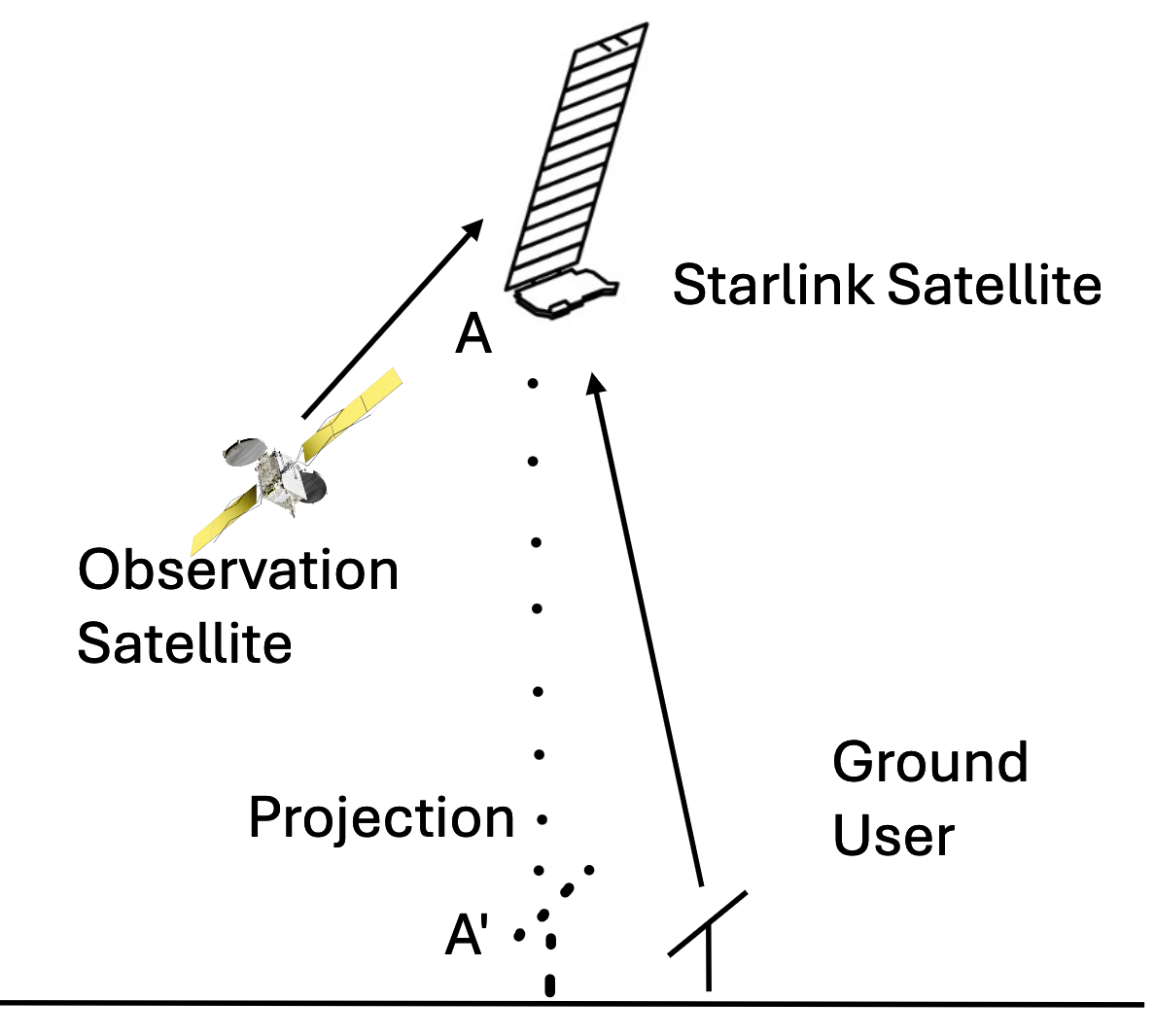}
    \vspace{-0.20in}
  \end{center}
  \caption{Space user to Starlink link estimation.}
  \vspace{-0.1in}
  \label{fig:Projection}
\end{wrapfigure}
To estimate the link rate between a space user and the satellite, we model the link based on a ground user who is close to the projection point of the space user. As illustrated in Figure~\ref{fig:Projection}, the Starlink satellite's position (A) is projected onto the Earth’s surface (A'), where we query the PoP profiling model under ideal weather conditions, approximating the space environment. The estimated link quality is then refined by incorporating: (1) ISL-induced loss and (2) additional attenuation accounting for differences in communication distance and atmospheric conditions between ground and space users.
Due to limited access to detailed satellite data, we do not directly model space-to-Starlink links. Instead, we reformulate the problem as estimating the incremental degradation caused by atmospheric effects. This is achieved via: (i) evaluating the sensitivity of our model to different weather inputs, and (ii) applying ITU-standard radio wave attenuation models that consider factors such as altitude, rain, cloud, and fog.

An implicit assumption is that a ground user connects to the satellite directly overhead, an assumption supported by scheduling data showing that over 80\% of Starlink’s selected satellites had elevation angles between 45° and 90°, despite only 30\% of all available satellites falling in this range~\cite{tanveer2023makingsenseconstellationsmethodologies}. Our projection-based method thus provides a practical and data-driven approximation of space-to-Starlink link quality, despite limited access to proprietary network details.

\subsection{Trace-driven Evaluation}
We utilize open-source orbital data of Starlink and observation satellites to emulate their movement. Each observation satellite is assumed to generate 500 GB of data daily \cite{295557}. Using the System Scheduler (Section 4.3), we schedule the links between observation and Starlink satellites. The quality of each route is estimated by combining the methods outlined in Sections 4.1 and 5.2. Finally, we calculate the total data transferred across all routes within the system.
The dataset of observation satellites is sourced from Planet Inc., featuring orbital information from 113 satellites within the Planet satellite constellation \cite{PlanetConstellation}. We incorporate traces based on the Planet constellation into our experiments. We model a scenario with three times the number of observation satellites, similar to those in the Planet constellation, competing for communication links. 
We collect TLEs which are NASA’s standard format of encoding space objects’ trajectories \cite{space-track} for all 7061 working Starlink satellites and calculate the position,and velocity of Starlink satellites in orbit using the PyOrbital library \cite{Pyorbital}. 


\begin{figure*}[t]
  \centering
  \begin{subfigure}[t]{0.32\textwidth}
    \centering
    \includegraphics[width=\linewidth]{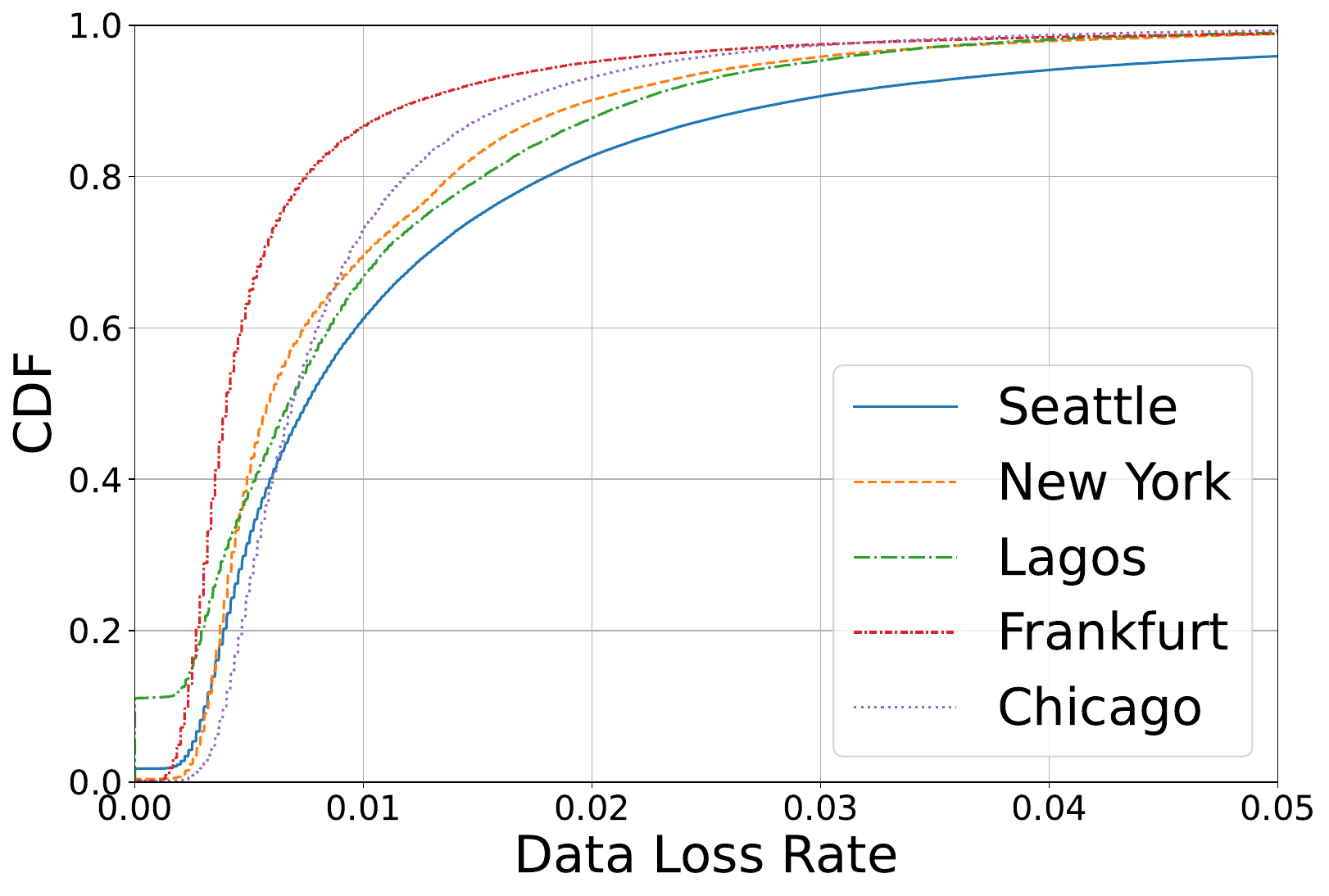}
    \caption{Packet loss rate for different PoPs.}
    \label{fig:PoPDataLossRateCDF}
  \end{subfigure}
  \hfill
  \begin{subfigure}[t]{0.32\textwidth}
    \centering
    \includegraphics[width=\linewidth]{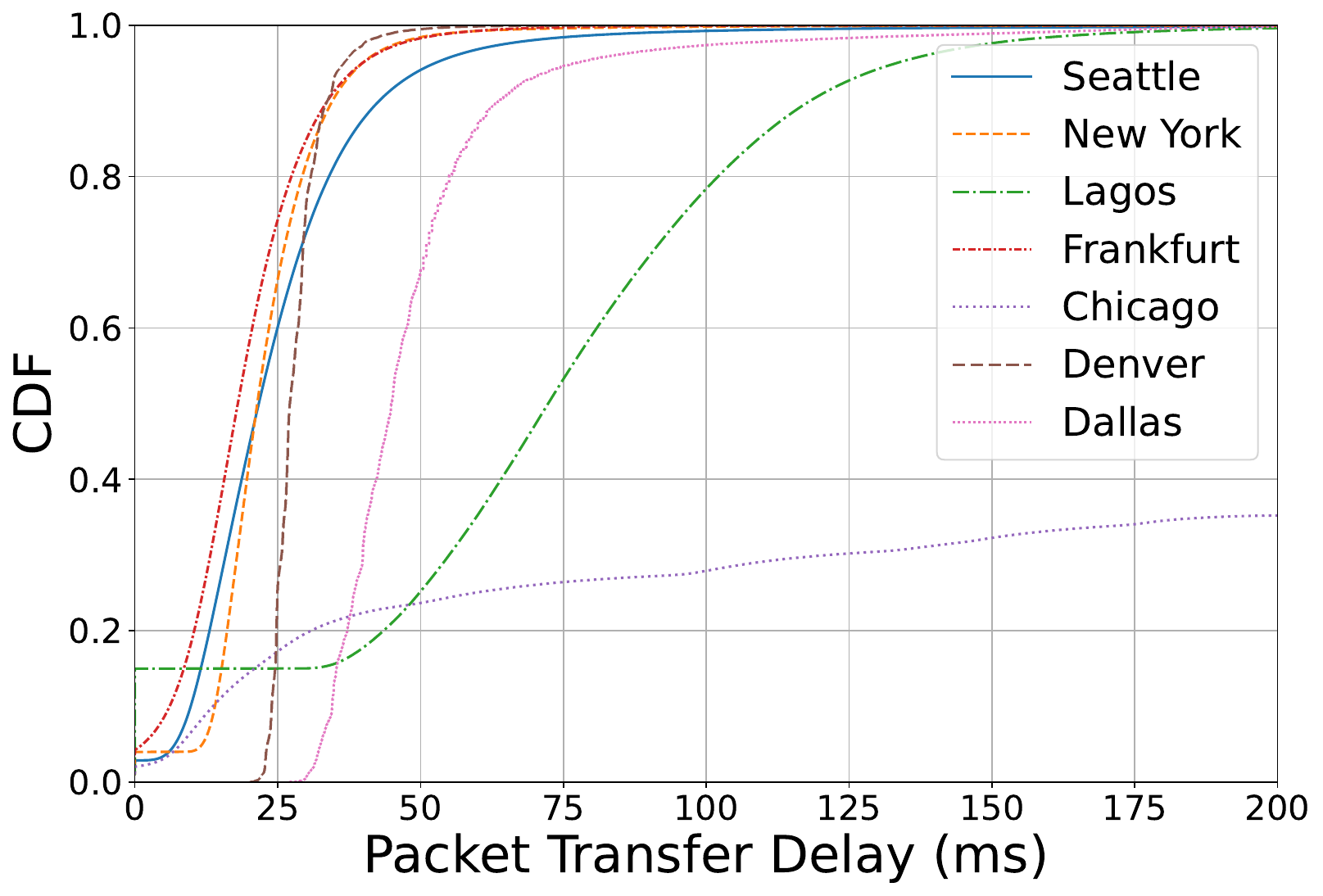}
    \caption{Packet delay for different PoPs.}
    \label{fig:DelayProfilePoPs}
  \end{subfigure}
  \hfill
  \begin{subfigure}[t]{0.32\textwidth}
    \centering
    \includegraphics[width=\linewidth]{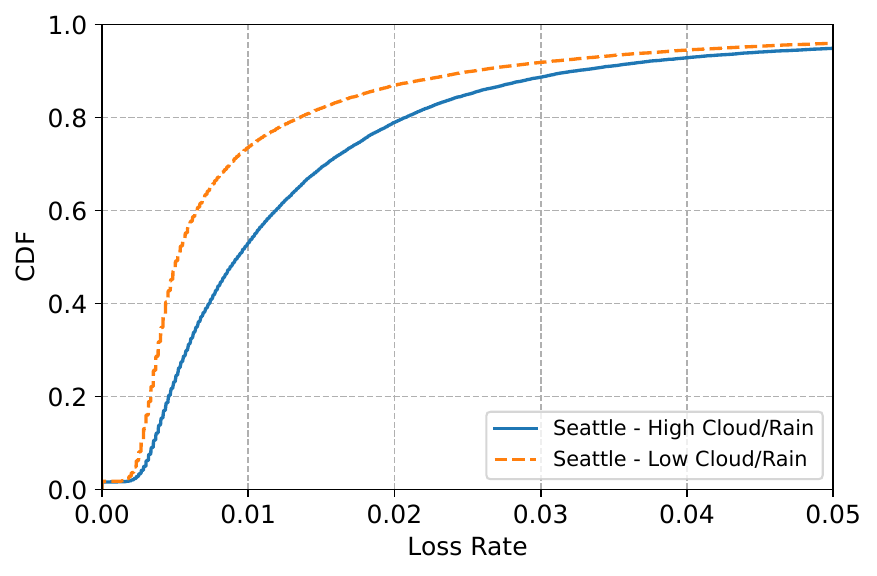}
    \caption{Packet loss for PoP at Seattle in different weather conditions.}
    \label{fig:LossSeattleWeather}
  \end{subfigure}
  \vspace{-0.15in}
  \caption{Comparison of packet loss and delay characteristics for different PoPs and weather conditions.}
  \vspace{-0.15in}
  \label{fig:PoPPerformanceCompare}
\end{figure*}

\subsection{Model Details}

We use 80\% of the dataset for training, leaving 20\% for testing. The dataset comprises four months of data, preprocessed to remove invalid entries and standardized before model training. We utilize a combination of XGBoost and a neural network, implemented in Keras \cite{Keras}, to model the loss rate. 
The training and evaluation processes are conducted on a cloud equipped with multiple GPUs using TensorFlow’s \texttt{MirroredStrategy}. The combined model, which integrates predictions from both XGBoost and the neural network, is further refined using a linear regression model. The single model training process completes in $\sim$4.5 hours with 4 Nvidia L40S, and predictions are made with milliseconds latency.

\subsection{Baselines and Models in Comparison}
Our primary baseline is the wait-and-transfer architecture utilizing high-throughput links as described in \cite{PlanetHighSpeedRadio}. This architecture employs six parallel channels alongside high-end receivers equipped with 4-meter diameter dish antennas, designed to maximize data transfer efficiency in a single session. For comparison with the setup in \cite{PlanetHighSpeedRadio}, which models five high-throughput ground stations globally, we simulate five Starlink PoPs distributed across the planet. In our evaluation of route quality estimation, we benchmark against a simple regression predictor.
We compare the system data transfer ability with L2D2 \cite{L2D2}, which deploys 173 low-cost ground stations for satellite data downlink.


\begin{table}
\centering
\small 
\setlength{\tabcolsep}{2pt} 
\begin{tabular}{|l|cc|cc|cc|cc|}
\hline
                    & \multicolumn{2}{c|}{\textbf{Standard}} & \multicolumn{2}{c|}{\textbf{Roam}} & \multicolumn{2}{c|}{\textbf{Priority}} & \multicolumn{2}{c|}{\textbf{Business}} \\ \hline
\textbf{Error/Kbps} & \textbf{SSU} & \textbf{RP} & \textbf{SSU} & \textbf{RP} & \textbf{SSU} & \textbf{RP} & \textbf{SSU} & \textbf{RP} \\ \hline
\textbf{Chicago}    & 34.6         & 76.8        & 46.1         & 102.4        & 76.0         & 169.0        & 138.2         & 307.2        \\
\textbf{Frankfurt}  & 38.4         & 99.8        & 51.2         & 133.1        & 84.5         & 219.7        & 153.6         & 399.4        \\
\textbf{New York}   & 39.9         & 115.2       & 53.2         & 153.6        & 87.9         & 253.4        & 159.7         & 460.8        \\
\textbf{Lagos}      & 63.0         & 138.2       & 84.0         & 184.3        & 138.5        & 304.1        & 251.9         & 553.0        \\
\textbf{Seattle}    & 46.8         & 92.2        & 62.5         & 122.9        & 103.1        & 202.8        & 187.4         & 368.6        \\ \hline
\end{tabular}
\caption{Data rate difference due to estimation error.}
\vspace{-0.2in}
\label{tab:DataRateError}
\end{table}

\section{RESULTS}

\subsection{PoP Profiling}
\begin{wrapfigure}{R}{0.3\textwidth}
  \begin{center}
    \vspace{-0.25in}
    \includegraphics[width=0.29\textwidth]{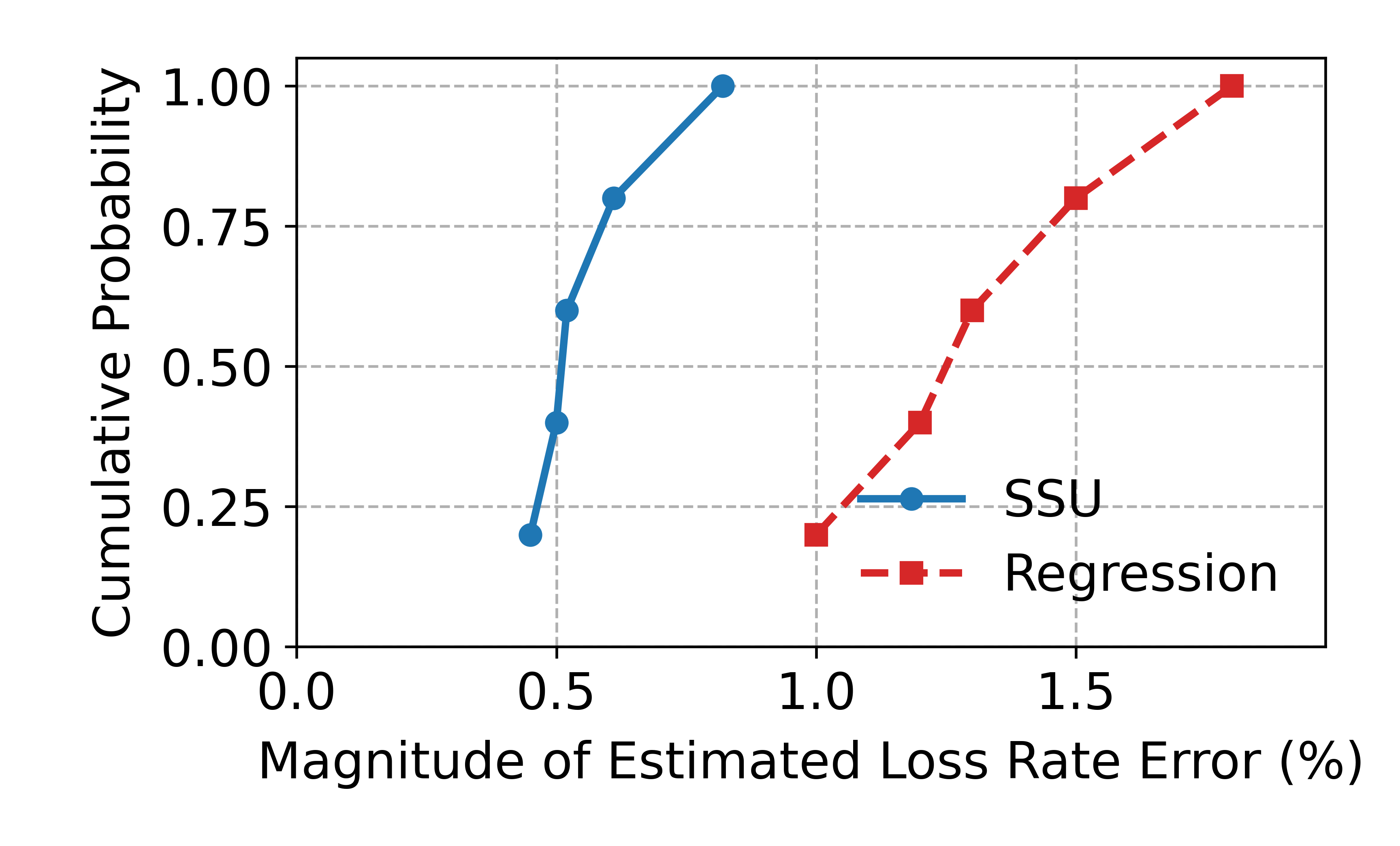}
    \vspace{-0.25in}
  \end{center}
  \caption{Loss rate prediction.}
  \vspace{-0.2in}
  \label{fig:EstimationErrorCompare}
\end{wrapfigure}
To assess the effectiveness of the PoP profiling and ISL estimation model, we utilize the real-world measurements from Starlink POPs presented in Table~\ref{tab:starlink_dishes}. Figure~\ref{fig:EstimationErrorCompare} shows the cumulative distribution function (CDF) of the mean absolute error between the predicted and observed packet loss rates across all PoPs. The SSU model achieves a median error of 0.52\% in terms of loss rate, with the error at the 90th percentile being 0.74\%. This small mean absolute error suggests that the data rate prediction is unlikely to deviate significantly between different configurations. In comparison, the regression predictor results in a median error of 1.3\% (with a 90th percentile error of 1.68\%), which is roughly twice as high. This discrepancy is likely due to the regression predictor's design, which focuses on statistical summaries without accounting for the granular variations caused by specific PoP characteristics and route components, such as detailed weather fluctuations.

To explore the impact of packet loss rate error on data rates, we convert the predicted packet loss rates by both SSU and regression predictor into their corresponding subscribed data rates according to the service tiers. Table~\ref{tab:DataRateError} reports the results. Two important observations arise from this table: firstly, SSU consistently outperforms the regression predictor by a considerable margin; secondly, some PoPs exhibit higher profiling errors than others, primarily due to limited data availability. For example, the PoP associated with the Seychelles dish was rerouted from Lagos to Frankfurt on Dec. 08, 2023, reducing the amount of usable data for Lagos. These results indicate that the SSU route estimation model provides a more accurate data rate prediction, highlighting its potential as a reliable foundation for SSU.

To gain insights into POP profiles, we examine two metrics that influence data transfer: packet loss rate and packet transfer delay. Figure~\ref{fig:PoPDataLossRateCDF} presents the CDF of mean packet loss rates for five PoPs in two months. New York and Frankfurt exhibit lower packet loss rates, indicating more reliable data transmission, whereas Lagos shows higher loss rates, emphasizing the impact of regional network stability.  Figure~\ref{fig:DelayProfilePoPs} plots the CDF of packet transfer delay across seven PoPs, showing the variation across different locations. Locations like Frankfurt and Seattle demonstrate lower delays compared to Lagos and Dallas, which have higher latency and greater variability. This highlights how geographical and infrastructural factors affect transfer performance.
Note that the performance of a PoP can vary significantly under different weather conditions. For instance, the elevated delay observed in Chicago may be attributed to increased cloud cover and snowfall during the measurement period (i.e., December and January). To further investigate weather-induced performance degradation, we compare the loss rates under clear conditions versus high cloud and rain scenarios. As shown in Figure~\ref{fig:LossSeattleWeather}, the loss rate of Seattle PoP under high cloud/rain conditions exhibits a relative increase of 36.44\%.

\begin{figure}
  \centering
  \includegraphics[width=0.7\linewidth]{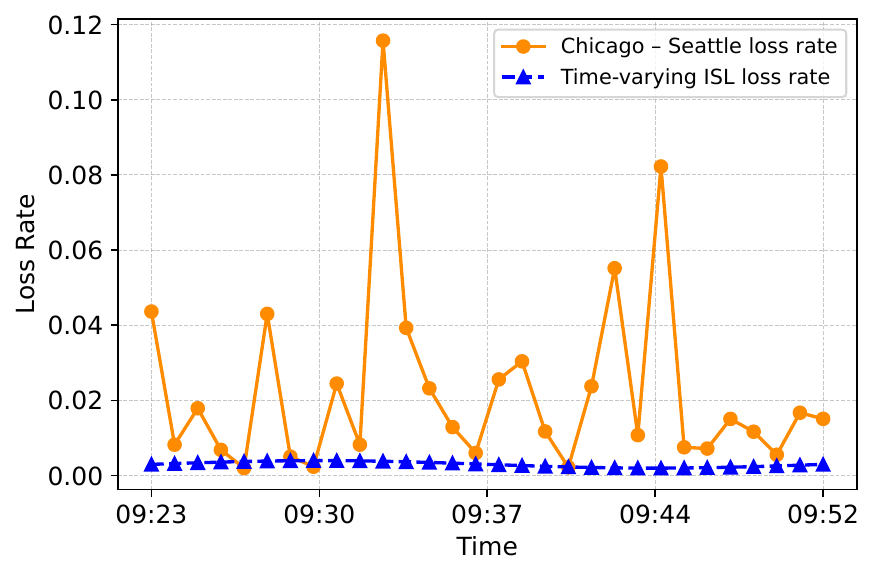}
  \vspace{-0.2in}
  \caption{An instance of loss rate variation between 2 PoPs.}
  \label{fig:LossChicagoCompareISL}
  \vspace{-0.05in}
\end{figure}

Although the average loss rate over the two-month period reported in Figure~\ref{fig:PoPDataLossRateCDF} appears relatively low, significant variations can occur within specific short-term intervals, influencing the selection of destination PoPs. For instance, Figure~\ref{fig:LossChicagoCompareISL} shows a scenario where, during a half-hour interval, the loss rate between Chicago and Seattle is sufficiently high so that the optimal choice of destination PoP would be Seattle, despite the space user’s proximity to Chicago.

\subsection{ISL Estimation}

\begin{figure}
  \centering
  \includegraphics[width=0.7\linewidth]{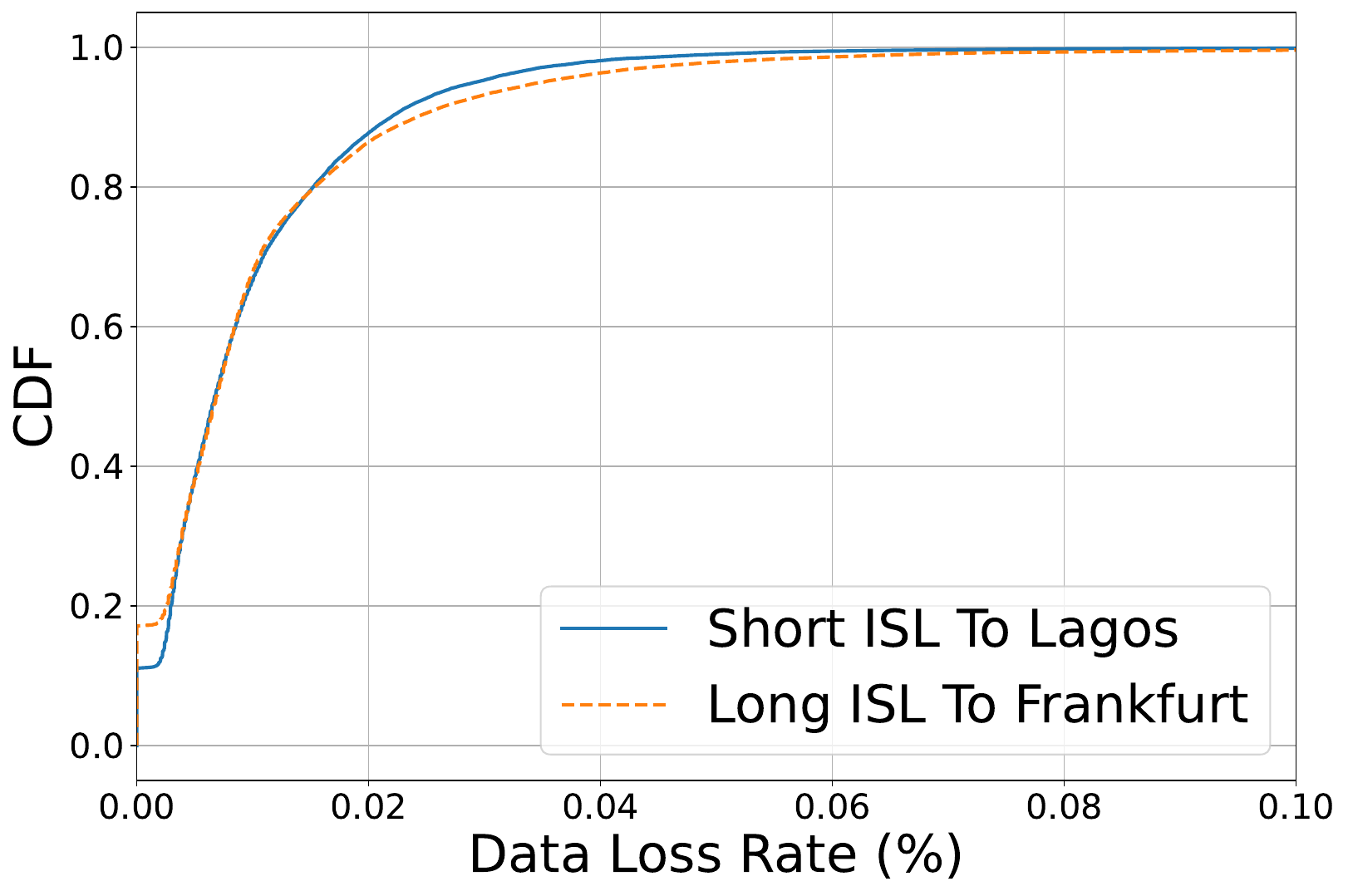}
  \vspace{-0.15in}
  \caption{Packet loss rate comparison of short and long ISL.}
  \label{fig:ISLcompareOneMonth}
  \vspace{-0.05in}
\end{figure}

As reported in Figure~\ref{fig:ISLcompareOneMonth}, the actual recorded loss rate difference we obtained between Lagos and Frankfurt observed from Seychelles via ISL has an average value of 0.08\%. By incorporating the estimated loss rate 0.27\%, which reflects the local conditions at the Frankfurt PoP, into this difference, it allows us to isolate the loss rate difference specifically attributable to the ISL connection between Lagos and Frankfurt observed from Seychelles. Over a span of 23 days, the mean loss rate difference caused by ISL discrepancies is 0.35\%. This difference reflects the additional degradation in link quality due to the extended path via ISL (2100km) from Seychelles to Frankfurt compare with Seychelles to Lagos.
The mean ISL-induced loss rate (0.35\%/2100km) serves as the reference for estimating the ISL-related loss rates for other users within our system. We assume that the increase in loss rate caused by ISL is proportional to the distance spanned by the ISL. 

\subsection{Link Selection Efficiency}
\begin{figure}
  \centering
  \begin{subfigure}[t]{0.48\linewidth}
    \centering
    \includegraphics[width=\linewidth]{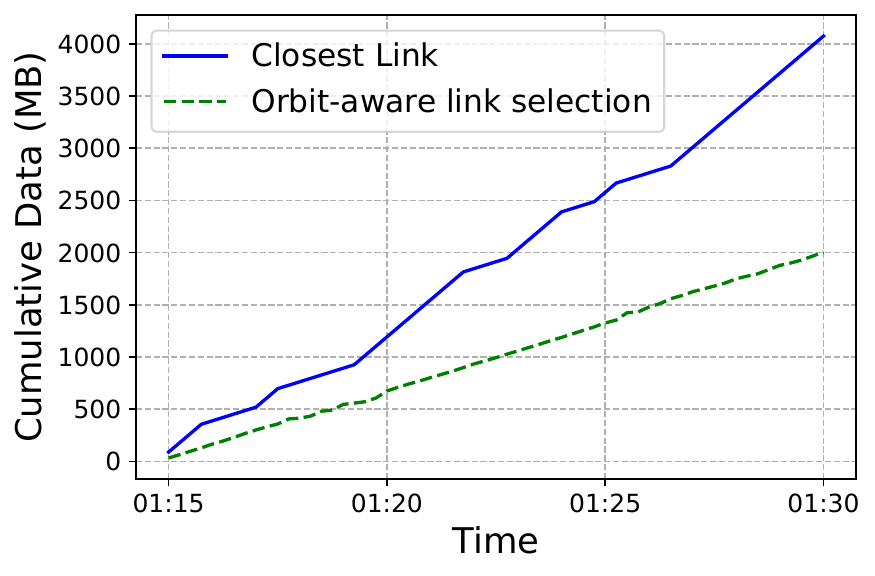}
    \vspace{-0.2in}
    \caption{Inside SSDB: closest vs ours.}
    \label{fig:CompareCuDataInSSDB}
  \end{subfigure}
  \hfill
  \begin{subfigure}[t]{0.48\linewidth}
    \centering
    \includegraphics[width=\linewidth]{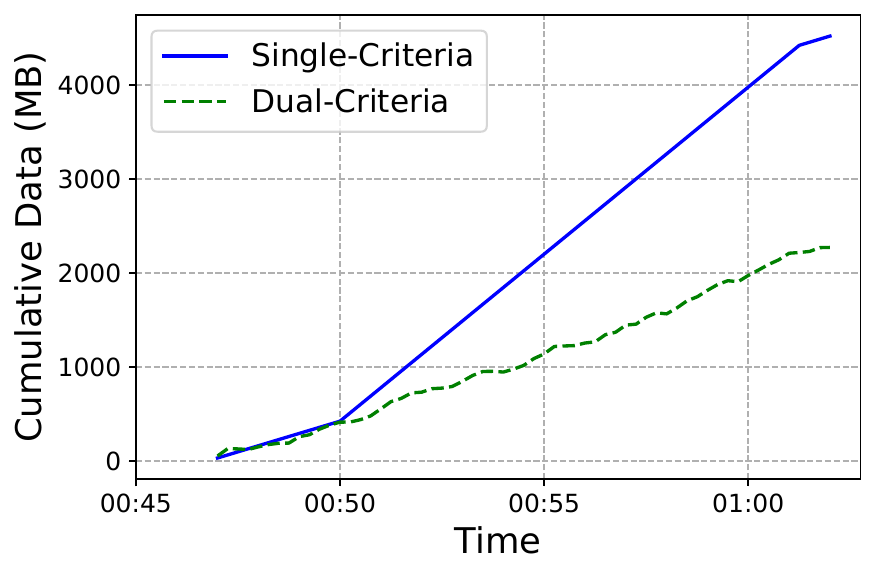}
    \vspace{-0.2in}
    \caption{Outside SSDB: single-criterion vs dual-criterion.}
    \label{fig:DataCumOutSSDB}
  \end{subfigure}
  \vspace{-0.1in}
  \caption{Comparing two link selection strategies.}
  \label{fig:SSDB_Comparison}
\end{figure}

To validate the efficiency of our link selection strategy within the SSDB, we compare it against a baseline approach that intuitively selects the link with the shortest distance. 
Under the same data production rate from the space users (500 GB/day), we compare the cumulative volume of data backlog that are not transferred. As shown in Figure~\ref{fig:CompareCuDataInSSDB}, our orbit-aware link selection strategy achieves nearly twice the efficiency of the shortest-distance baseline.
Outside the SSDB, we compare our dual-criterion strategy against a single-criteria selection strategy that consistently chooses satellites with relative velocities close to $v_G$. As shown in the Figure~\ref{fig:DataCumOutSSDB}, this approach remains competitive with SSU near the SSDB boundary, where satellites with velocities close to $v_G$ are still available. However, beyond this region, its performance rapidly degrades, and our proposed dual-criteria selection strategy achieves significantly more data transfer.

\subsection{System Data Transfer Ability}
We utilize orbital information from 113 satellites within the Planet constellation \cite{PlanetConstellation} to evaluate the data downlink transfer ability of SSU. We assume there are three times as many satellites competing for downlink around each Planet observation satellite. We compare four variants: SSU, SSU (75\%), SSU(Polar Outage) and SSU-random. In SSU, we assume Starlink for business service tier (30 Mbps) is used and fewer ground users are utilizing the Starlink facilities, providing ample connection opportunities for all observation satellites to maintain continuous connectivity. SSU (75\%) simulates a congestion scenario where competition arises between ground and space users, resulting in an average of 25\% of space users losing connectivity to the Starlink system. SSU(Polar Outage) stands for performance when no similar polar orbit can be found and connection gets outage. The SSU-random selects a random Starlink satellite from the ones selected via the orbit-aware link selection strategy (Section 4.2), serving to demonstrate the effectiveness of the scheduler in Section 4.3.

\begin{figure}
  \centering
  \includegraphics[width=0.9\linewidth]{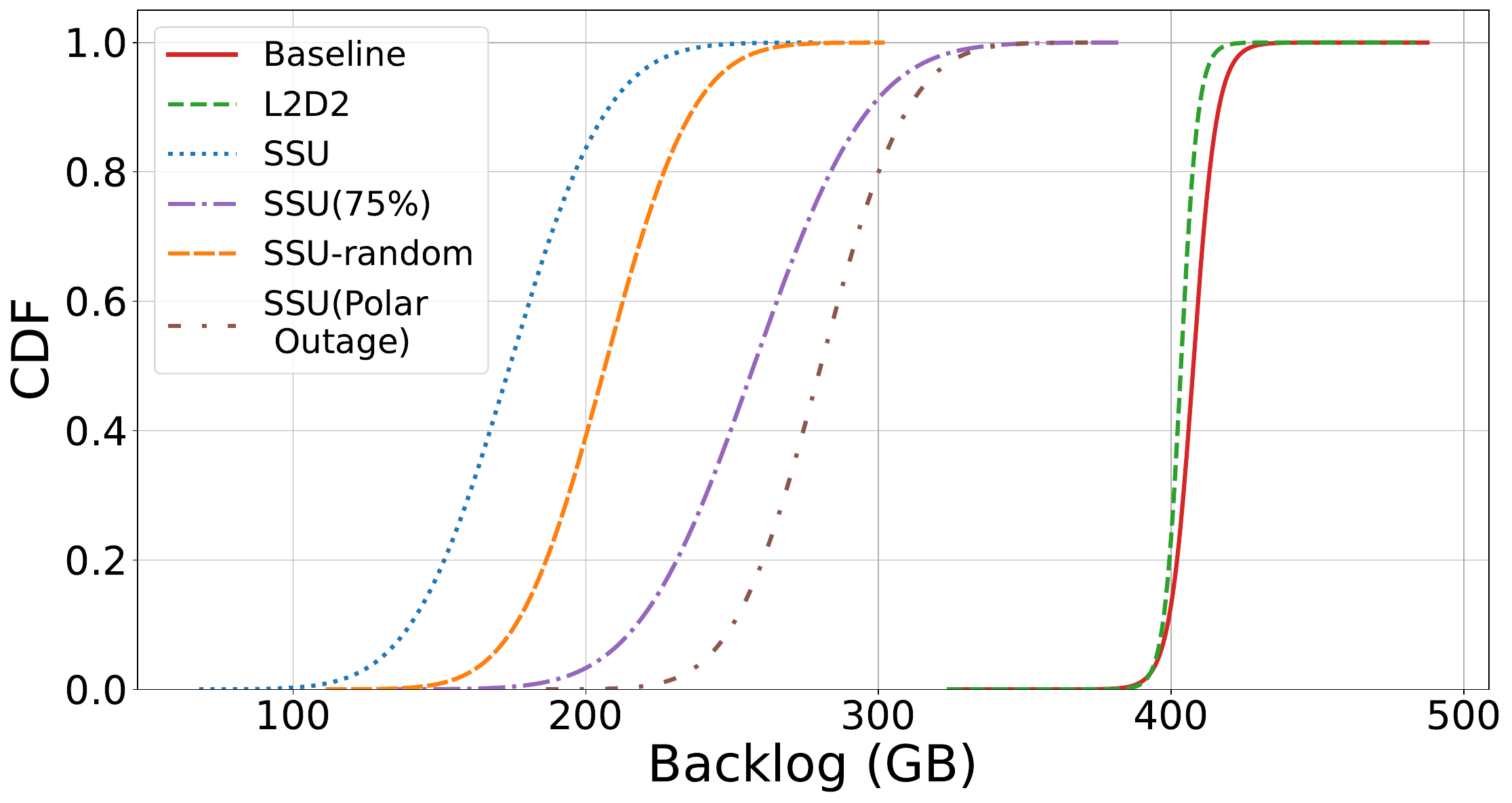}
  \vspace{-0.15in}
  \caption{Data backlog for one observation satellite.}
  \label{fig:BacklogForOneDay}
\end{figure}

We measure the amount of data remaining to be downloaded from observation satellites at the end of the day and plot the CDF in Figure~\ref{fig:BacklogForOneDay}. For the baseline scenario, the median backlog is 407.6 GB, with a 90th-percentile backlog of 426.5 GB, indicating that 10\% of the satellites still have 426.5 GB of data pending download. In comparison, the SSU variant demonstrates a significantly reduced backlog, with a median of 174.0 GB and a 90th-percentile backlog of 208.0 GB, which also outperforms the random scheduler (207.6/238.2 GB) by more than 10\%. Even under the congestion conditions simulated in SSU (75\%) or SSU (Polar Outage), the system achieves a backlog of 257.4/280.3 GB, which represents only 63.2\% (68.7\%) of the baseline values. The performance surpasses that of L2D2 \cite{L2D2} by more than three times as well.

\subsection{Cost Analysis}
Since Earth observation satellites would act as \textit{space users} of the Starlink network, consumer-grade Local Priority plans are not applicable. We therefore evaluate two relevant service tiers: (i) the Global Priority plan at \$1,150/TB (1~TB) and \$1,075/TB (2~TB) \cite{starlinkplans2024}; and (ii) the Business Unlimited (Aviation) plan at \$10,000/month for unlimited usage \cite{starlinkaviation2024}. The effective cost under the latter depends on sustainable throughput: ranging from about \$772/TB at 40~Mbps down to \$77/TB at 400~Mbps. Incorporating amortized Ku-band hardware cost (about \$150k over 5 years \cite{starlinkhardware2024}) adds approximately \$190/TB at 40~Mbps to only \$19/TB at 400~Mbps. 

Overall, Starlink Global Priority produces unit costs comparable to the traditional ground-station model (exemplified by Planet Labs):\$1,075--\$1,150/TB, while the Business Unlimited plan—if accessible for satellites—can reduce costs to below \$200/TB under sustained throughputs above 200~Mbps. These results suggest that a Starlink-based relay has the potential to deliver substantially lower cost per TB than conventional ground station architectures.

\section{RELATED WORK}
 \textbf{Data Downlink for Observation Satellites:}
Prior studies address satellite data downlink from various angles. Traditional "wait-and-transfer" strategies optimize downlink within limited contact windows \cite{devaraj2017dove, devaraj2019planet}, but remain constrained by spectrum, hardware, and energy limitations. Other efforts enhance ground infrastructure, either by deploying portable ground stations \cite{riesing2017portable} or expanding terrestrial coverage through dedicated facilities \cite{L2D2, martin2018amazon}. To mitigate congestion, some works apply network congestion control to improve throughput \cite{10.1145/3570361.3592521}, while industry solutions (e.g., AWS, Azure, LeafSpace) rent ground station access to increase availability \cite{aws_ground_station, azure_orbital, leafspace_leafline}. Edge computing has also been explored to prioritize or preprocess data \cite{denby2020orbital, 295557}, though it introduces on-orbit complexity. In contrast, our approach leverages LEO satellite communication without specialized infrastructure, aiming for full data delivery and improved throughput via constellation-level scheduling.

\textbf{Starlink Measurement and Analysis:}
With Starlink’s expansion, researchers have increasingly studied its performance. Early studies assessed throughput, latency, and reliability under protocols like TCP and QUIC \cite{michel2022first}, and showed advantages over conventional networks \cite{kassem2022browser}. Further analyses revealed beam-switching dynamics \cite{garcia2023multi} and global handover intervals \cite{tanveer2023making}. The emergence of inter-satellite links (ISLs) was evidenced in \cite{chaudhry2021laser}, followed by large-scale measurement across user terminals \cite{zhao2024lens, hu2023leo, izhikevich2024democratizing, raman2023dissecting}. Additionally, satellite mobility and routing behavior were explored in \cite{li2023networking}. Distinct from prior work, we focus on routing-level performance and ISL quality by building ML-based PoP profiles and link estimators, offering a new lens to understand Starlink’s end-to-end data transfer capabilities.

\textbf{Orbit Computing:} Another approach to improving data transfer for EO satellites is \emph{orbit computing} \cite{bhattacherjee2020orbit}. This method leverages the increasing onboard computing power of satellites to pre-process, filter, and select useful data before transmission, thereby reducing the total volume of data that must be downlinked. Such a concept and prior work \cite{295557} are not contradictory to our method; rather, they are complementary and can be combined to further enhance data delivery efficiency.

\section{CONCLUSION}
In this paper, we present SSU, a data transfer framework that leverages Starlink infrastructure to enable low-latency data delivery from observation satellites. We evaluate its feasibility through extensive data analysis and introduce three key innovations: Starlink PoP profiling, orbit-aware link selection, and system-level scheduling optimization. Real-world measurements and large-scale trace-driven evaluations confirm the superior performance of our design.

\balance
\bibliographystyle{ACM-Reference-Format}
\bibliography{satellite_research}

\appendix

\end{document}